\documentstyle[12pt]{article}

\def\thefootnote{\fnsymbol{footnote}}

\newlength{\minitwocolumn}
\catcode`\@=11
\long\def\@makefntext#1{
\protect\noindent \hbox to 3.2pt {\hskip-.9pt  
$^{{\eightrm\@thefnmark}}$\hfil}#1\hfill}               

\def\thefootnote{\fnsymbol{footnote}}
\def\@makefnmark{\hbox to 0pt{$^{\@thefnmark}$\hss}}    
        
\def\ps@myheadings{\let\@mkboth\@gobbletwo
\def\@oddhead{\hbox{}
\rightmark\hfil\eightrm\thepage}   
\def\@oddfoot{}\def\@evenhead{\eightrm\thepage\hfil
\leftmark\hbox{}}\def\@evenfoot{}
\def\sectionmark##1{}\def\subsectionmark##1{}}

\font\eightrm=cmr8

\def\MPL{Mod.~Phys.~Lett. }

\def\PL{Phys.~Lett. }
\def\PR{Phys.~Rev. }

\def\PTP{Prog.~Theor.~Phys. }

\def\brs{\delta}

\newcommand{\rd}{\overleftarrow{\partial}} 
\newcommand{\ld}{\overrightarrow{\partial}} 

\newcommand{\sbv}[2]{{({{#1},{#2}})}}

\def\bV{\mbox{\boldmath $V$}}
\def\bX{\mbox{\boldmath $X$}}
\def\bY{\mbox{\boldmath $Y$}}
\def\bZ{\mbox{\boldmath $Z$}}

\def\ba{\mbox{\boldmath $A$}}
\def\bb{\mbox{\boldmath $B$}}

\def\bR{\mbox{\boldmath $R$}}

\def\bphi{\mbox{\boldmath $\phi$}}

\newcommand{\calA}{{\cal A}}
\newcommand{\calB}{{\cal B}}
\newcommand{\calC}{{\cal C}}
\newcommand{\calD}{{\cal D}}

\newcommand{\calI}{{\cal I}}
\newcommand{\calJ}{{\cal J}}
\newcommand{\calL}{{\cal L}}

\newcommand{\aone}{{\mbox{\boldmath $\tilde{A_1}{}$}}}
\newcommand{\azero}{{\mbox{\boldmath $\tilde{\alpha_0}{}$}}}
\newcommand{\btwo}{{\mbox{\boldmath $\tilde{B_2}{}$}}}
\newcommand{\bone}{{\mbox{\boldmath $\tilde{B_1}{}$}}}
\newcommand{\beone}{{\mbox{\boldmath $\tilde{\beta_1}{}$}}}
\newcommand{\bezero}{{\mbox{\boldmath $\tilde{\beta_0}{}$}}}
\newcommand{\phizero}{{\mbox{\boldmath $\tilde{\phi}{}$}}}
\newcommand{\phiminus}{{\mbox{\boldmath $\tilde{\phi}_{-1}$}}}
\newcommand{\dtwo}{{d}{}}
\newcommand{\aonep}{{\mbox{\boldmath $\tilde{A_1^{\prime}}{}$}}}
\newcommand{\azerop}{{\mbox{\boldmath $\tilde{\alpha_0^{\prime}}{}$}}}
\newcommand{\btwop}{{\mbox{\boldmath $\tilde{B_2^{\prime}}{}$}}}
\newcommand{\bonep}{{\mbox{\boldmath $\tilde{B_1^{\prime}}{}$}}}
\newcommand{\beonep}{{\mbox{\boldmath $\tilde{\beta_1^{\prime}}{}$}}}
\newcommand{\bezerop}{{\mbox{\boldmath $\tilde{\beta_0^{\prime}}{}$}}}

\newcommand{\phiminusp}{{\mbox{\boldmath $\tilde{\phi}_{-1}^{\prime}$}}}

\newcommand{\aonesub}{{\mbox{$\tilde{A_1}{}$}}}
\newcommand{\azerosub}{{\mbox{$\tilde{\alpha_0}{}$}}}
\newcommand{\btwosub}{{\mbox{$\tilde{B_2}{}$}}}
\newcommand{\bonesub}{{\mbox{$\tilde{B_1}{}$}}}
\newcommand{\beonesub}{{\mbox{$\tilde{\beta_1}{}$}}}
\newcommand{\bezerosub}{{\mbox{$\tilde{\beta_0}{}$}}}
\newcommand{\phizerosub}{{\mbox{$\tilde{\phi}{}$}}}

\newcommand{\Q}{{\kern.24em\vrule width.04em height1.4ex%
                 depth-.05ex\kern-.26em\mathsf Q}}
\newcommand{\C}{{\kern.24em\vrule width.04em height1.4ex%
                 depth-.05ex\kern-.26em\mathsf C}}

\newcommand{\Sprime}{{\Sigma^{\prime}}}

\begin{document}




\begin{titlepage}
\begin{flushright}
YITP-04-15
\end{flushright}

\vskip 1.35cm
\begin{center}
{\Large \bf
An Alternative Topological Field Theory \\ of Generalized Complex Geometry
}
\vskip 1.2cm
Noriaki IKEDA$^1$%
\footnote{E-mail address:\ ikeda@yukawa.kyoto-u.ac.jp}
\ and Tatsuya TOKUNAGA$^2$
\footnote{E-mail address:\ tokunaga@yukawa.kyoto-u.ac.jp}
\vskip 0.4cm
{\it $^1$Department of Mathematical Sciences,
Ritsumeikan University \\
Kusatsu, Shiga 525-8577, Japan }\\
{\it $^2$Yukawa Institute for Theoretical Physics,
Kyoto University \\
Kyoto 606-8502, Japan }\\
\vskip 1.5cm

\today

\begin{abstract}
We propose a new topological field theory on generalized complex geometry
 in two dimension 
 using AKSZ formulation.
  Zucchini's model is $A$ model in the case that the generalized complex structure depends on only a symplectic structure.  
Our new model is $B$ model in the case that the generalized complex structure depends on only a complex structure.  
%
\end{abstract}
\end{center}
\end{titlepage}

\renewcommand{\thefootnote}{\alph{footnote}}

\setcounter{page}{2}


\rm
\section{Introduction}
\noindent
In \cite{Zucchini:2004ta}\cite{Zucchini:2005rh}, Zucchini has 
constructed a two dimensional topological sigma model on generalized 
complex geometry \cite{GHR} \cite{Hit} \cite{Gua} by the AKSZ formulation \cite{Alexandrov:1995kv} (also see \cite{Roytenberg:2006qz}), 
which is a general geometrical framework to construct 
a topological sigma model by the Batalin-Vilkovisky formalism \cite{Bat}.  
Also, there are many recent papers \cite{Kap}-\cite{Lindstrom:2007xv} 
on this topic.  
Zucchini's model is a generalization of the 
Poisson sigma model and is similar to A model in \cite{Alexandrov:1995kv}.  
However B model looks different from the Zucchini model 
because B model has more fields than the Zucchini model has.  

In this paper, we propose an alternative realization
of generalized complex geometry by a topological field theory 
by the AKSZ formulation.  
Our model is similar to B model, not A model
in the sense of AKSZ, as a worldsheet action of a topological 
sigma model with superifields on a supermaifold. 
Our model is the first candidate which naturally includes B model 
and may be related to a topological string theory on 
generalized Calabi-Yau geometry \cite{Chuang:2006vt} \cite{Zucchini:2006ii}.  
%

First we construct a three dimensional topological field theory 
of generalized complex geometry with a nontrivial $3$-form $H$, which has Zucchini's model as a boundary action.  
This topological field theory is a reconstruction by the AKSZ formulation 
of the model proposed in the paper \cite{Ikeda:2004cm}.
%
%
%
%
%
%
%
%
Next after a dimensional reduction, 
we derive a topological field theory of generalized complex geometry 
in two dimensions from three dimensions.  
We can see that this model has a generalized complex structure as a consistency condition of a topological BV action.  
If the generalized complex structure is a complex structure, 
our model has one parameter marginal deformation of the model 
without changing a complex structure, 
and reduces to B model in a limit of the deformation.
If the generalized complex structure is a symplectic structure, 
our model becomes a new 2D topological sigma model with a symplectic structure.

The paper is organized as follows. 
In section 2, the AKSZ actions of A model, B model and the Zucchini model 
are reviewed.
In section 3, three dimensional topological field theory
of generalized complex geometry is rederived in the AKSZ formulation.  
In section 4, we derive a two dimensional topological field 
theory of generalized complex geometry and check its properties.  
In section 5, our model is reduced in two special ways.  
Section 6 includes conclusion and discussion.  
In appendix A, a generalized complex structure is briefly summarized.
In appendix B, the AKSZ formulation of the Batalin-Vilkovisky 
formalism in general $n$ dimensions is reviewed.  

\section{$A$ Model, $B$ Model and Zucchini Model}
\noindent
In this section, we review the 
AKSZ formulation of topological sigma models such as 
$A$ model, $B$ model and the Zucchini model.  
\subsection{$A$ Model and $B$ Model}
\noindent
$A$ model and $B$ model are defined on 
the graded bundle
\begin{eqnarray}
T^*[1] M \oplus \left(T[1]M \oplus T^*[0]M \right).
\label{2dtmspace}
\end{eqnarray}
Here $E = TM$, $n=2$ and $p \geq 1$ 
in the general graded bundles (\ref{totspaceike}).  
Local coordinates are written by superfields on this bundle:
$(\bphi^i, \bb_{1i}, \ba_1{}^{i}, \bb_{0,i})$.  
$\bphi^i$ is a map $\bphi^i:\Pi T \Sigma \rightarrow M$,
and $\bb_{1i}$ is a basis of sections of $\Pi T^* \Sigma \otimes \bphi^*(T^* [1] M)$.  
$\ba_1{}^{i}$ is a basis of sections of $\Pi T^* \Sigma \otimes \bphi^*(T [1] M)$, and 
$\bb_{0i}$ is a basis of sections of $\Pi T^* \Sigma \otimes \bphi^*(T^* [0] M)$.  
%
%
%
%
The antibracket on this bundle (\ref{2dtmspace}) is 
\begin{eqnarray}
\sbv{F}{G} & \equiv &
F \frac{\rd}{\partial \bphi^i} 
\frac{\ld }{\partial \bb_{1,i}}  G
- 
F \frac{\rd }{\partial \bb_{1,i}} 
\frac{\ld }{\partial \bphi^i} G
+ F  \frac{\rd}{\partial \ba_1{}^{i}} 
\frac{\ld }{\partial \bb_{0,i}}  G
-
F \frac{\rd }{\partial \bb_{0,i}} 
\frac{\ld }{\partial \ba_1{}^{i}}  G
\label{2Dantibracket}
\end{eqnarray}
from (\ref{bfantibracketike}).

The $A$ model action with a symplectic form 
$Q_{ij}$ in \cite{Witten:1988xj} is 
\begin{eqnarray}
S_{AQ} = \frac{1}{2} \int_{\Pi T \Sigma} 
Q_{ij}(\bphi) d \bphi^i d \bphi^j,
\label{oaaction}
\end{eqnarray}
where $d$ is a superderivative $d = \theta^{\mu} \partial_{\mu}$.  
where the integration $\int_{\Pi T \Sigma}$ means
the integration on the supermanifold, 
$\int_{\Pi T \Sigma} d^2 \theta d^2 \sigma$.
This action is consistent if and only if 
the $2$-form $Q = \frac{1}{2} Q_{ij} d \phi^i d \phi^j$ satisfies the symplectic condition $d_M Q = 0$, namely
\begin{eqnarray}
\partial_k Q_{ij} + \partial_i Q_{jk} + \partial_j Q_{ki} = 0.  
\label{Qcondition}
\end{eqnarray}
This model is rewritten by the AKSZ formulation 
on the graded bundle 
$
T^*[1] M \oplus \left(T[1]M \oplus T^*[0]M \right) 
$.  
We introduce $\ba_1^i$, $\bb_{0i}$ and $\bb_{1i}$ 
as auxiliary fields, and rewrite the action using the first
order formalism.  
The action in AKSZ formulation is 
\begin{eqnarray}
S_{AQ} = \int_{\Pi T \Sigma} 
\left( \bb_{1i} d \bphi^i - \bb_{0i} d \ba_1^i 
- \bb_{1i} \ba_1^i 
+ \frac{1}{2} Q_{ij}(\bphi) \ba_1^i \ba_1^j
\right).
\label{aqmodel}
\end{eqnarray}
We can check that $\sbv{S_{AQ}}{S_{AQ}} = 0$ if and only if 
the $2$-form $Q$ satisfies the symplectic condition (\ref{Qcondition}).
%
%
%

Also, $A$ model action with a Poisson bivector $P^{ij}$ 
is 
\begin{eqnarray}
S_{AP} = \int_{\Pi T \Sigma}
\bb_{1i} d \bphi^i - \bb_{0i} d \ba_1^i 
+ \frac{1}{2} P^{ij}(\bphi) \bb_{1i} \bb_{1j}, 
\label{apmodel}
\end{eqnarray}
which is called the Poisson sigma model
\cite{Ikeda:1993aj}\cite{Schaller:1994es}.  
The consistency condition $\sbv{S_{AP}}{S_{AP}} = 0$ is satisfied if and only if 
$P^{ij}$ is a Poisson bivector field i.e.
\begin{eqnarray}
P^{il} \partial_l P^{jk} 
+ P^{jl} \partial_l P^{ki} + P^{kl} \partial_l P^{ij} = 0.
\label{Poissoncondition}
\end{eqnarray}

$B$ model with a complex structure $J^i{}_j$ is 
\begin{eqnarray}
S_B = \int_{\Pi T \Sigma}
\bb_{1i} d \bphi^i - \bb_{0i} d \ba_1^i 
+ J^i{}_j(\bphi) \bb_{1i} \ba_1^j
+ \frac{\partial J^i{}_k}{\partial \bphi^j}(\bphi) 
\bb_{0i} \ba_1^j \ba_1^k,
\label{bmodel}
\end{eqnarray}
which is a covariant form of $B$ model action in \cite{Alexandrov:1995kv},
but 
is different from the action in \cite{Hofman:2002cw}.  
%
We can check that the consistency condition $\sbv{S_{B}}{S_{B}} = 0$ is satisfied if and only if 
$J^i{}_j$ satisfies the integrability condition for the complex 
structure
\begin{eqnarray}
J^l{}_i \partial_l J^k{}_j - J^l{}_j \partial_l J^k{}_i
- J^k{}_l \partial_i J^l{}_j + J^k{}_l \partial_j J^l{}_i = 0.
\end{eqnarray}

\subsection{Zucchini Model}
\noindent
In \cite{Zucchini:2004ta}, Zucchini has proposed a topological sigma model 
with a generalized complex structure on 
a two dimensional worldsheet $\Sigma$.  
Although he called this model ''the Hitchin sigma model'', 
here we call it the Zucchini model.

First we consider $H = 0$ case.
The action of the Zucchini's model is 
\begin{eqnarray}
S_Z = \int_{\Pi T \Sigma}
\bb_{1i} d \bphi^i 
+ \frac{1}{2} P^{ij}(\bphi) \bb_{1i} \bb_{1j} 
+ \frac{1}{2} Q_{ij}(\bphi) d \bphi^i d \bphi^j
+ J^i{}_j(\bphi) \bb_{1i} d \bphi^j.
\label{zucchini}
\end{eqnarray}
The master equation
$\sbv{S_Z}{S_Z} =0$ is satisfied if $P$, $Q$ and $J$ satisfy
the conditions for a generalized complex structure (\ref{J2zero}), 
(\ref{PQantisym}) and (\ref{intcoord}).  
We can see that the Batalin-Vilkovisky structure of this model defines 
a generalized complex structure on a target manifold $M$.  
If $J^i{}_j = 0$ in the action (\ref{zucchini}), 
the action reduces to the summation of two realizations of $A$ model such that (\ref{oaaction}) + (\ref{apmodel}).  
%
However, if $P^{ij} = Q_{ij} = 0$, the action (\ref{zucchini}) 
does not reduce to the $B$ model action (\ref{bmodel}).  
So we can not easily see whether the 
Zucchini model can be related to $B$ model.

Also, we can consider $b$-transformation property of this model \cite{Zucchini:2004ta}.  
%
The $b$-transformation is defined by (\ref{btrans}), (\ref{btransj}) and 
\begin{eqnarray}
&& \hat{\bphi}^{i} = \bphi^i, 
\nonumber \\
&& \hat{\bb}_{1i} = \bb_{1i} + b_{ij} d \bphi^j.
\label{2dbtrans}
\end{eqnarray}
The $b$-transformation produces the $b$ field term such as 
\begin{eqnarray}
\hat{S}_Z = S_Z - \int_{\Pi T \Sigma} b_{ij} d \bphi^i d \bphi^j.  
\label{btzucchini}
\end{eqnarray}
This suggests that the Zucchini action with $H \neq 0$
should have a Wess-Zumino term 
\begin{eqnarray}
S_{ZH} = \int_{\Pi T \Sigma}
\bb_{1i} d \bphi^i 
+ \frac{1}{2} P^{ij} \bb_{1i} \bb_{1j} 
+ \frac{1}{2} Q_{ij} d \bphi^i d \bphi^j 
+ J^i{}_j \bb_{1i} d \bphi^j
+ \frac{1}{2} \int_{\Pi T X}
H_{ijk} d \bphi^i d \bphi^j d \bphi^k,
\label{ztzucchini}
\end{eqnarray}
where
$X$ is a three dimensional worldvolume such that 
$\Sigma = \partial X$ is a two dimensional boundary of $X$.


\section{3D Topological Field Theory with Generalized Complex Structures
from 2D Zucchini Model}
\noindent
In this section, we review a three dimensional topological field theory 
with a generalized complex structure 
from the Zucchini model in two dimensions.  Here this topological field theory is redefined by the AKSZ formulation, which was not explicitly written in \cite{Ikeda:2004cm}.  
%
%

\subsection{$H=0$ case}
\noindent
Let $X$ be a three dimensional worldvolume with a coordinate $(\sigma^M)$ for $M = 1, 2, 3$,
and $\Sigma = \partial X$ be a two dimensional boundary of $X$.  
First we consider $H = 0$ case.  
%

By using the Stokes theorem, we can see the action (\ref{zucchini}) as 
\begin{eqnarray}
S_Z &=& \int_{\Pi T X}
d \left( \bb_{1i} d \bphi^i 
+ \frac{1}{2} P^{ij} \bb_{1i} \bb_{1j} 
+ \frac{1}{2} Q_{ij} d \bphi^i d \bphi^j 
+ J^i{}_j \bb_{1i} d \bphi^j \right)
\nonumber \\
&=& \int_{\Pi T X}
d \bb_{1i} d \bphi^i 
+ \frac{1}{2} \frac{\partial P^{ij}}{\partial \bphi^k} 
d \bphi^k \bb_{1i} \bb_{1j} 
+ P^{ij} d \bb_{1i} \bb_{1j} 
+ \frac{1}{2} \frac{\partial Q_{ij}}{\partial \bphi^k} 
d \bphi^k d \bphi^i d \bphi^j 
\nonumber \\
&&
+ \frac{\partial J^i{}_j}{\partial \bphi^k} d \bphi^k \bb_{1i} d \bphi^j 
+ J^i{}_j d \bb_{1i} d \bphi^j,
\label{boundary3D}
\end{eqnarray}
where $d$ is a three dimensional derivative $d = \theta^M \partial_M$.  
$\bphi^i$ and $\bb_{1i}$ can be extended to those on $X$ such that 
$\bphi^i:\Pi T X \rightarrow M$
and $\bb_{1i}$ is a basis 
of sections of $\Pi T^* X \otimes \bphi^*(T^* [1] M)$.  
We introduce a superfield $\ba_1^i$ with total degree one,
which is a basis of a section of 
$\Pi T^* X \otimes \bphi^*(T [1] M)$ 
such that $\ba_1^i = d \bphi^i$, 
and
a superfield $\bb_{2i}$ with total degree two,
which is a basis of a section of 
$\Pi T^* X \otimes \bphi^*(T^* [2] M)$ 
such that $\bb_{2i} = - d \bb_{1i}$.  
Moreover, we 
introduce two Lagrange multiplier fields $\bY_{2i}$ and $\bZ_1^i$ 
in order to realize two equations such as 
$\ba_1^i = d \bphi^i$ and
$\bb_{2i} = - d \bb_{1i}$
by the equations of motion.  
The superfield $\bY_{2i}$ with total degree two
is 
a section of $\Pi T^* X \otimes \bphi^*(T^* [2] M)$,
and the superfield $\bZ_1^i$ with total degree one
is 
a section of $\Pi T^* X \otimes \bphi^*(T [1] M)$.  
The 3D action (\ref{boundary3D}) is equivalent to 
\begin{eqnarray}
S_Z &=& \int_{\Pi T X}
- \bb_{2i} \ba_1^i 
+ \frac{1}{2} \frac{\partial P^{ij}}{\partial \bphi^k} 
\ba_1^k \bb_{1i} \bb_{1j} 
- P^{ij} \bb_{2i} \bb_{1j} 
+ \frac{1}{2} \frac{\partial Q_{jk}}{\partial \bphi^i} \ba_1^i \ba_1^j \ba_1^k 
\nonumber \\
&&
+ \frac{\partial J^i{}_j}{\partial \bphi^k} \ba_1^k \bb_{1i} \ba_1^j 
- J^i{}_j \bb_{2i} \ba_1^j
+ (\ba_1^i - d \bphi^i ) \bY_{2i}
+ (\bb_{2i} + d \bb_{1i} ) \bZ_1^i.
\label{baction}
\end{eqnarray}
We define $\bY^\prime_{2i} = \bY_{2i} - \frac{1}{2} \bb_{2i}$
and $\bZ_1^{\prime i} = \bZ_1^i - \frac{1}{2} \ba_1^i$.  
The action (\ref{baction}) is rewritten as 
\begin{eqnarray}
&& S_Z = S_a + S_{b} + \mbox{total derivative} ~ ;
\nonumber \\
&& S_{a} =  
\int_{\Pi T X} - \bY^\prime_{2i} d \bphi^i + d \bb_{1i} \bZ_1^{\prime i}
+ \bY^\prime_{2i} \ba_1^i + \bb_{2i} \bZ_1^{\prime i},
\nonumber \\
&& S_{b}
= \int_{\Pi T X} - \frac{1}{2} \bb_{2i} d \bphi^i 
+ \frac{1}{2} \bb_{1i} d \ba_1^i 
- J^i{}_j \bb_{2i} \ba_1^j
- P^{ij} \bb_{2i} \bb_{1j} 
+ \frac{1}{2} 
\frac{\partial Q_{jk}}{\partial \bphi^{i}}
\ba_1^i \ba_1^j \ba_1^k
\nonumber \\ 
&& \quad \quad \ 
+  \frac{1}{2} \left(
- \frac{\partial J^k{}_{j}}{\partial \bphi^i} 
+ \frac{\partial J^k{}_{i}}{\partial \bphi^j} \right)
\ba_1^i \ba_1^j \bb_{1k}
+  \frac{1}{2} 
\frac{\partial P^{jk}}{\partial \bphi^i} 
\ba_1^i \bb_{1j} \bb_{1k}.
\label{boundarysigma}
\end{eqnarray}
where $S_a$ is independent of a generalized complex structure.
$S_b$ can be written as 
\begin{eqnarray}
S_b &=& \int_{\Pi T X} 
- \frac{1}{2} \langle 0 + \bb_{2}, d ( \bphi + 0) \rangle
+ \frac{1}{4} \langle \ba_1 + \bb_1, d (\ba_1 + \bb_1) \rangle
\nonumber \\
&& - \langle 0 + \bb_{2}, \calJ  (\ba_1 + \bb_{1}) \rangle
- \frac{1}{2} \langle  \ba_1 + \bb_1, 
\ba_1^i \frac{\partial \calJ}{\partial \bphi^i}
(\ba_1 + \bb_1) \rangle
+ \mbox{total derivative},
\end{eqnarray}
which is analogical with the $B$ model action (\ref{bmodel}).

The antibracket (P-structure) on $X$, which is 
induced from the antibracket (\ref{2Dantibracket}) on $\Sigma$, for $\bphi^i$, $\bb_{2,i}$, $\ba_1{}^{i}$ and 
$\bb_{1,i}$ 
is given by the antibracket (\ref{bfantibracketike}) in $n=3$.  
In order to define the antibrackets for $\bY^{\prime}_{2i}$ 
and $\bZ_1^{\prime i}$, 
we introduce two antibracket conjugate fields $\bX^i$,
which are maps from $\Pi T X$ to $M$,
and $\bV_{1i}$, 
which are sections of $\Pi T^* X \otimes \bphi^*(T^* [1] M)$.  
The model is defined on the graded bundle of the 
direct product of $T^*[2] M \oplus \left(T[1]M \oplus T^*[1]M \right)$
and $(T[0]M \oplus T^*[2] M) \oplus \left(T[1]M \oplus T^*[1]M \right)$.  
The second bundle is represented by auxiliary fields.  
The antibracket is 
\begin{eqnarray}
\sbv{F}{G} & \equiv &
F \frac{\rd}{\partial \bphi^i} 
\frac{\ld }{\partial \bb_{2,i}}  G
- 
F \frac{\rd }{\partial \bb_{2,i}} 
\frac{\ld }{\partial \bphi^i} G
+ F  \frac{\rd}{\partial \ba_1{}^{i}} 
\frac{\ld }{\partial \bb_{1,i}}  G
+
F \frac{\rd }{\partial \bb_{1,i}} 
\frac{\ld }{\partial \ba_1{}^{i}}  G
\nonumber \\
&&
+ F \frac{\rd}{\partial \bX^i} 
\frac{\ld }{\partial \bY_{2,i}^{\prime}}  G
- 
F \frac{\rd }{\partial \bY_{2,i}^{\prime}} 
\frac{\ld }{\partial \bX^i} G
+ F  \frac{\rd}{\partial \bZ_1{}^{\prime i}} 
\frac{\ld }{\partial \bV_{1,i}}  G
+
F \frac{\rd }{\partial \bV_{1,i}} 
\frac{\ld }{\partial \bZ_1{}^{\prime i}}  G.
\label{gcsab}
\end{eqnarray}
%
We can check that
$S_Z$ satisfies the master equation $\sbv{S_Z}{S_Z} = 0$ 
if $J$, $P$ and $Q$ are components of the generalized complex 
structure (\ref{gcstcomp}).  
We can take the proper boundary conditions $\Sigma = \partial X$;
\begin{equation}
{\ba_1{}^i}_{//}|_{\partial X} = 0, 
{\bb_{2i}}_{//}|_{\partial X} = 0, 
{\bY_{2i}^{\prime}}_{//}|_{\partial X} = 0,
{\bZ_1{}^{\prime i}}_{//}|_{\partial X} = 0,
\end{equation}
such that the total derivative 
terms on the master equation $\sbv{S_Z}{S_Z}$ vanish.  Here ${//}$ means 
that we take the components which are tangent to the boundary $\partial X$.

Also, because $\sbv{S_a}{S_a} = \sbv{S_a}{S_b} = 0$,
$S_b$ satisfies the master equation $\sbv{S_b}{S_b} = 0$ 
\begin{eqnarray}
&& \calA{}^{ijk} = \calB{}_i{}^{jk} = \calC{}_{ij}{}^k = 0,
\nonumber \\
&& \partial_i \calD{}_{jkl}
+ (ijkl \;\; \rm{cyclic}) = 0,
\label{cond}
\end{eqnarray}  
where $\calA{}^{ijk}$, $\calB{}_i{}^{jk}$, $\calC{}_{ij}{}^k$ and $\calD{}_{jkl}$ are defined in Appendix A.  
Therefore, 
we can see $S_{b}$ as a three dimensional AKSZ action 
with generalized complex structure.  
We discuss why the condition is 
not $\calD{}_{jkl} = 0$ but
$ \partial_i \calD{}_{jkl}
+ (ijkl \;\; \rm{cyclic}) = 0$ 
in subsection 3.3.

%
%
%
We call $S_b$ three dimensional 
generalized complex sigma model.

We consider 3D $b$-transformation property from the 2D $b$-transformations (\ref{2dbtrans}) and the conditions
$\ba_1^i = d \bphi$ and ${\bb}_{2i} = - d \bb_{1i}$.   
3D $b$-transformations are 
\begin{eqnarray}
&& \hat{\bphi}^i = \bphi^i,
\nonumber \\
&& \hat{\ba_1}^i = \ba_1^i,
\nonumber \\
&& \hat{\bb}_{1i} = \bb_{1i} + b_{ij} \ba_1^j,
\nonumber \\
&& \hat{\bb}_{2i} = {\bb}_{2i} - d ( b_{ij} \ba_1^j),
\nonumber \\
&& \hat{\bY}^{\prime}_{2i} = {\bY}^{\prime}_{2i} 
+ \frac{1}{2} b_{ij} d \ba_1^j 
- \frac{1}{2} \frac{\partial b_{ik}}{\partial \bphi^j} \ba_1^j d \bphi^k 
\nonumber \\ 
&& \qquad
- J^l{}_k \frac{\partial b_{jl}}{\partial \bphi^i} \ba_1^j \ba_1^k 
- P^{lk} \frac{\partial b_{jl}}{\partial \bphi^i} \ba_1^j \bb_{1k}
+ d (J^l{}_k b_{li} \ba_1^k )
+ d (P^{lk} b_{li} \bb_{1k}), 
\nonumber \\
&& \hat{\bZ_1}^{\prime i} = \bZ_1^{\prime i}.
\label{btransfield}
\end{eqnarray}
%
We can see that 3D action (\ref{boundarysigma}) is invariant under the $b$-transformation such that 
%
\begin{eqnarray}
\hat{S}_Z = S_Z .
\label{btransbtzucchini}
\end{eqnarray}
%

\subsection{$H \neq 0$ case I~:Action induced from the Zucchini model}
\noindent
In the similar way, we can consider the case of a twisted generalized complex structure with $H \neq 0$.  
~From the Zucchini model with $H \neq 0$ (\ref{ztzucchini}), 
a three dimensional action is derived as 
\begin{eqnarray}
&& S_{ZH} = S_{a} + S_{Hb} + \mbox{total derivative} ~ ;
\nonumber \\
&& S_{a} = 
\int_{\Pi T X} - \bY^\prime_{2i} d \bphi^i + d \bb_{1i} \bZ_1^{\prime i}
+ \bY^\prime_{2i} \ba^i + \bb_{2i} \bZ_1^{\prime i},
\nonumber \\
&&
S_{Hb} =  \int_{\Pi T X} 
- \frac{1}{2} \bb_{2i} d \bphi^i + \frac{1}{2} \bb_{1i} d \ba_1^i 
- J^i{}_j \bb_{2i} \ba_1^j
- P^{ij} \bb_{2i} \bb_{1j} 
+ \frac{1}{2} 
\left( H_{ijk} + \frac{\partial Q_{jk}}{\partial \bphi^{i}}
\right) \ba_1^i \ba_1^j \ba_1^k
\nonumber \\ 
&& \quad \quad \ 
+  \frac{1}{2} \left(
- \frac{\partial J^k{}_{j}}{\partial \bphi^i} 
+ \frac{\partial J^k{}_{i}}{\partial \bphi^j} \right)
\ba_1^i \ba_1^j \bb_{1k}
+  \frac{1}{2} 
\frac{\partial P^{jk}}{\partial \bphi^i} 
\ba_1^i \bb_{1j} \bb_{1k}.
\label{ztboundarysigma}
\end{eqnarray}
This action (\ref{ztboundarysigma}) 
satisfies the master equation $\sbv{S_{ZH}}{S_{ZH}} = 0$, 
if $J$, $P$, $Q$ and $H$ are components of 
a twisted generalized complex structure (\ref{tgcstcomp}).  
However, this action is not $b$-invariant under the $b$-transformation 
(\ref{btransfield}), (\ref{btrans}) and (\ref{btransj}).
The action (\ref{ztboundarysigma}) transforms under the $b$-transformation as
\begin{eqnarray}
\hat{S}_{ZH} 
= S_{ZH} 
- \int_{\Pi T X} 
\frac{3}{2} \frac{\partial b_{jk}}{\partial \bphi^i} 
\ba_1^i \ba_1^j \ba_1^k
= S_{ZH} 
- { 1 \over2} \int_{\Pi T X} 
(d_M b)_{[ijk]}  \ba_1^i \ba_1^j \ba_1^k,
\label{btransbtzucchini}
\end{eqnarray}
which has been expected from $b$-transformation property
 (\ref{btzucchini}) in the two dimensional model.

Since $H$ is closed, from the Poincar\'e Lemma, 
we can {\it locally} write $H$ 
with a $2$-form $q$ 
on $M$ such as
\begin{eqnarray}
H_{ijk} = \frac{1}{2} 
\left( \frac{\partial q_{jk}}{\partial \bphi^i} 
+ \frac{\partial q_{ki}}{\partial \bphi^j} 
+ \frac{\partial q_{ij}}{\partial \bphi^k} \right).
\end{eqnarray}
The $\frac{\partial b_{jk}}{\partial \phi^i} \ba_1^i \ba_1^j \ba_1^k$ term in
(\ref{ztboundarysigma})
can be absorbed to $Q$ by a 
{\it local} $b$-transformation $q_{ij} = b_{ij}$ 
in the action (\ref{ztboundarysigma}), 
and we obtain just the $H=0$ action (\ref{boundarysigma}).  
In other words, the $H$ terms in (\ref{ztboundarysigma}) are consistent 
up to $H$-exact terms as a global theory,
 and this model is meaningful only as a cohomology class in $H^3(M)$.  
It is a gerbe gauge transformation dependence 
\cite{Zucchini:2004ta}.

%
%
%
%

If we set $Q_{ij} = J^i{}_j = 0$ in  (\ref{ztzucchini}),
we obtain the AKSZ formulation of 
the WZ-Poisson sigma model \cite{KS}:
\begin{eqnarray}
S_{WZP} &=& \int_{\Pi T \Sigma}
\bb_{1i} d \bphi^i 
+ \frac{1}{2} P^{ij} \bb_{1i} \bb_{1j}
+ \frac{1}{2} \int_{X}
H_{jkl} d \bphi^i d \bphi^j d \bphi^k.
\label{WZpoisson}
\end{eqnarray}
~From (\ref{ztboundarysigma}), 
the 3D topological sigma model equivalent to (\ref{WZpoisson}) is 
\begin{eqnarray}
S_{WZP} &=& S_a + S_{WZPb} ~ ;
\nonumber \\
S_a &=&  
\int_{\Pi T X} - \bY^\prime_{2i} d \bphi^i + d \bb_{1i} \bZ_1^{\prime i}
+ \bY^\prime_{2i} \ba_1^i + \bb_{2i} \bZ_1^{\prime i},
\nonumber \\
S_{WZPb}
&=& \int_{\Pi T X} 
- \frac{1}{2} \bb_{2i} d \bphi^i + \frac{1}{2} d \bb_{1i} \ba_1^i 
- P^{ij} \bb_{2i} \bb_{1j} 
+ \frac{1}{2} 
H_{ijk}
\ba_1^i \ba_1^j \ba_1^k
\nonumber \\ 
&& \quad \quad \ 
+  \frac{1}{2} 
\frac{\partial P^{jk}}{\partial \bphi^i} 
\ba_1^i \bb_{1j} \bb_{1k}.
\label{boundaryWZpoisson}
\end{eqnarray}

\subsection{$H \neq 0$ case II~: $b$-invariant action}
\noindent
We can construct a $b$-invariant action with 
$H \neq 0$ in three dimensional manifold $X$.  
We introduce other $H$ terms.  
\begin{eqnarray}
&& S_I = S_{a} + S_{Ib} ~ ;
\nonumber \\
&& S_a =  
\int_{\Pi T X} 
- \bY^\prime_{2i} d \bphi^i + d \bb_{1i} \bZ_1^{\prime i}
+ \bY^\prime_{2i} \ba_1^i + \bb_{2i} \bZ_1^{\prime i},
\nonumber \\
&&
S_{Ib} 
=  \int_{\Pi T X} 
- \frac{1}{2} \bb_{2i} d \bphi^i + \frac{1}{2} \bb_{1i} d \ba_1^i 
- J^i{}_j \bb_{2i} \ba_1^j
- P^{ij} \bb_{2i} \bb_{1j} 
+ \frac{1}{2} 
\left( J^l{}_{i} H_{jkl} + \frac{\partial Q_{jk}}{\partial \bphi^{i}}
\right) \ba_1^i \ba_1^j \ba_1^k
\nonumber \\ 
&& \quad \quad \ 
+  \frac{1}{2} \left(- P^{kl} H_{ijl} 
- \frac{\partial J^k{}_{j}}{\partial \bphi^i} 
+ \frac{\partial J^k{}_{i}}{\partial \bphi^j} \right)
\ba_1^i \ba_1^j \bb_{1k}
+  \frac{1}{2} 
\frac{\partial P^{jk}}{\partial \bphi^i} 
\ba_1^i \bb_{1j} \bb_{1k}.
\label{tboundarysigma}
\end{eqnarray}
$S_{I}$ satisfies the master equation $\sbv{S_{I}}{S_{I}} = 0$ 
under the antibracket (\ref{gcsab})
if and only if $J$, $P$, $Q$ and $H$ are components of 
a twisted generalized complex structure.  
%
%
%
Namely, the master equation $\sbv{S_{I}}{S_{I}} = 0$ gives 
\begin{eqnarray}
&& \calA_H{}^{ijk} = \calB_H{}_i{}^{jk} = \calC_H{}_{ij}{}^k = 0,
\nonumber \\
&& \partial_i \calD_H{}_{jkl} + (ijkl \;\; \rm{cyclic}) = 0,
\label{condh}
\end{eqnarray}
where $\calA_H{}^{ijk}$, $\calB_H{}_i{}^{jk}$, $\calC_H{}_{ij}{}^k$ and $\calD_H{}_{jkl}$ are defined in Appendix A.  
%
%
The integrability condition is
not $\calD_H{}_{ijk} = 0$ 
but $\partial_i \calD_H{}_{jkl}
+ (ijkl \; \rm{cyclic}) = 0$ 
%
because 
the action $S_{I}$ is $b$-transformation invariant, 
$H{}_{ijk}$ has $b$-transformation ambiguity
by (\ref{btransj}), and $H$ is defined as a cohomology class in $H^3(M)$
in a twisted generalized complex structure.  

Since $S_{Ia}$ does not depend on a twisted generalized complex structure, 
$\sbv{S_{Ib}}{S_{Ib}} = 0$ is satisfied under the condition
(\ref{condh}).  
We can introduce the coupling constants by redefining 
${\bY}^{\prime}_{2i}$ and ${\bZ_1}^{\prime i}$
to $g_1 {\bY}^{\prime}_{2i}$ and $g_2 \bZ_1^{\prime i}$.  
%
If we take the limits that $g_1 \rightarrow 0$ and $g_2 \rightarrow 0$, 
then $S_I \rightarrow S_{Ib}$ and 
a twisted generalized complex structure does not change.  
We call this model $S_{Ib}$ a three dimensional twisted 
generalized complex sigma model.

We can change the $b$-transformation so that the action ${S_{I}}$ is invariant, though the action (\ref{tboundarysigma}) is 
not invariant under the original $b$-transformation (\ref{btransfield}).   
%
The $b$-transformations for $\bb_{2i}$
and ${\bY}^{\prime}_{2i}$ are changed to
\begin{eqnarray}
&& \hat{\bb}_{2i} = \bb_{2i} 
- \frac{1}{2} \frac{\partial b_{jk}}{\partial \bphi^i} 
\ba_1^j \ba_1^k,
\nonumber \\
&& \hat{\bY}^{\prime}_{2i} = {\bY}^{\prime}_{2i}
+ \frac{\partial b_{jk}}{\partial \bphi^i} 
\ba_1^j \bZ_1^k
- b_{ij} d \bZ_1^j,\label{newbtransfield}
\end{eqnarray}
%
%
and $b$-transformations for the other fields are the same 
as (\ref{btransfield}).  
Then we can check $\hat{S}_{I} = S_{I}$
after short calculation.



%

\section{2D Topological Field Theory of Generalized Complex Geometry}
\noindent
In this section, we propose a new two dimensional topological field theory 
of generalized complex geometry using the 3D topological field theory.  
First, only a part of the 3D BV formalism action is dimensionally reducted to in two dimension, and next this is modified in the 2D BV formalism such that the master equations determine just generalized complex structures.  
One important reason to have to take this unusual way is that generally, master equations of BV formalisms are not kept by a dimensional reduction.  

\subsection{$H=0$}
First we consider the $H=0$ case.  
%
%
%
We consider a dimensional reduction, which can keep a generalized complex structure, from a three dimensional 
worldvolume $X$ to a two dimensional manifold $\Sprime$.  
%
$X$ is compactified to $\Sprime \times S^1$.  
Then $\Pi T X$ is compactified to $\Pi T \Sprime \times \Pi T S^1$.  
It should be noticed that $\Sprime$ is generally a different manifold from $\Sigma$.  

Here, we take $X = \Sigma \times \bR^+$,
where $\Sigma$ has a local coordinate 
$(\sigma^1, \sigma^2)$ and $\bR^+ = [0, \infty)$ has
a local coordinate $(\sigma^3)$.  
The second component $(\sigma^2)$ is compactified such that $\Sprime = L \times \bR^+$, whose local coordinate is
$(\sigma^1, \sigma^3)$, 
where $L$ is a manifold in one dimension.
%
%
%
%
%
%
%
%
We formulate the 
dimensional reduction from 
a general three dimensional manifold $X$ to 
a general two dimensional manifold $\Sprime$.
Here we ignore Kaluza-Klein modes
and consider only massless sectors, because we will see that the consistent BV action can be constructed in two dimension even if these KK modes are omitted.  
It is not our purpose that we derive the two dimensional model which is completely equivalent to the 3D topological field theory.  
The target graded bundle for the three dimensional model, $T^*[2] M \oplus \left(T[1]M \oplus T^*[1]M \right)$, 
reduces to the graded bundle for the two dimensional model, 
$\left(T^*[1] M \oplus \left(T[-1]M \oplus T^*[2]M \right)\right)
\oplus 
\left( \left(T[0]M \oplus T^*[1]M \right) 
\oplus \left(T[1]M \oplus T^*[0]M \right) \right)$.  
Under the dimensional reduction $(\sigma^1, \sigma^2, \sigma^3) \rightarrow (\sigma^1, \sigma^3)$, 
the fields are reduced as follows.  
\begin{eqnarray}
&& \bphi^i(\sigma^1, \sigma^2, \sigma^3) 
= \phizero^i(\sigma^1, \sigma^3)
+ \theta^2 \phiminus^i(\sigma^1, \sigma^3),  \nonumber \\
&& \ba_1^i(\sigma^1, \sigma^2, \sigma^3) 
= \aone^i(\sigma^1, \sigma^3) 
+ \theta^2 \azero^i(\sigma^1, \sigma^3),  \nonumber \\
&& \bb_{1i}(\sigma^1, \sigma^2, \sigma^3) 
= \bone_i(\sigma^1, \sigma^3) 
+ \theta^2 \bezero_i(\sigma^1, \sigma^3),  \nonumber \\
&& \bb_{2i}(\sigma^1, \sigma^2, \sigma^3) 
= \btwo_{i}(\sigma^1, \sigma^3) 
+ \theta^2 \beone_{i}(\sigma^1, \sigma^3),
\label{fielddimred}
\end{eqnarray}
where $\phiminus^i$ has the total degree $-1$, 
$\phizero^i$, $\azero^i$ and $\bezero_i$ have the total degree $0$, 
$\aone^a$, $\bone_i$ and $\beone_{i}$ have the total degree $1$, and 
$\btwo_{i}$ has the total degree $2$.  
All these superfields do not depend on $\theta^2$.  
%
%
%

The antibracket induced from three dimensions is
\begin{eqnarray}
\sbv{F}{G} & \equiv &
F \frac{\rd}{\partial \phizero^i} 
\frac{\ld }{\partial \beone_{i}}  G
- 
F \frac{\rd }{\partial \beone_{i}} 
\frac{\ld }{\partial \phizero^i} G
+ 
F \frac{\rd}{\partial \phiminus^i} 
\frac{\ld }{\partial \btwo_{i}}  G
- 
F \frac{\rd }{\partial \btwo_{i}} 
\frac{\ld }{\partial \phiminus^i} G
\nonumber \\
&&
+ F \frac{\rd }{\partial \bezero_{i}} 
\frac{\ld }{\partial \aone^i} G
- F \frac{\rd}{\partial \aone^i} 
\frac{\ld }{\partial \bezero_{i}}  G
+ 
F \frac{\rd}{\partial \azero^i} 
\frac{\ld }{\partial \bone_{i}}  G
- 
F \frac{\rd }{\partial \bone_{i}} 
\frac{\ld }{\partial \azero^i} G.
\label{reductionantibracket}
\end{eqnarray}
%
%
%
%
%
We take a three dimensional AKSZ action $S_b$ (\ref{boundarysigma}) 
with a generalized complex structure.  
%
%
The existence of the negative total degree superfield $\phiminus^i$ 
complexifies the dimensional reduction in the AKSZ formulation.  
Generally in \cite{Ikeda:2004gp}, it is known that even if we substitute (\ref{fielddimred}) to 
(\ref{boundarysigma}), we do not obtain the correct 
AKSZ action in two dimensions, and we need 
more $\phiminus^i$ terms.  

In order to derive the correct AKSZ action, 
first we should consider the dimensional reduction via the non-BV formalism.  
The superfields are expanded by the ghost numbers to 
\begin{eqnarray}
&& \bphi^i = \phi^{(0)i} + \phi^{(-1)i}
+ \phi^{(-2)i}
+ \phi^{(-3)i},
\nonumber \\ 
&& 
\bb_{1i} = B_{1,i}^{(1)} + B_{1,i}^{(0)}
+ B_{1,i}^{(-1)} + B_{1,i}^{(-2)},
\nonumber \\
&& 
\ba_1{}^{i} = A_1^{(1)i}
+ A_{1}^{(0)i} + A_{1}^{(-1)i} 
+ A_{1}^{(-2)i},
\nonumber 
\\ 
&& 
\bb_{2,i} = 
B_{2,i}^{(2)}
+ B_{2,i}^{(1)}
+ B_{2,i}^{(0)} 
+ B_{2,i}^{(-1)},
\end{eqnarray}
where $\phi^{(-1)i} \equiv \theta^{M} \phi_{M}^{(-1)i}$, etc.  
After setting all the antifield with negative ghost numbers to zero, 
the following non-BV action is 
\begin{eqnarray}
&& S_{b}^{(0)}
= \int_{\Pi T X} - \frac{1}{2} B_{2i}^{(0)} d \phi^{(0)i}
+ \frac{1}{2} B_{1i}^{(0)} d A_1^{(0)i} 
- J^i{}_j B_{2i}^{(0)} A_1^{(0)j}
- P^{ij} B_{2i}^{(0)} B_{1j}^{(0)}
\nonumber \\ 
&& \quad \quad \ 
+ \frac{1}{2} 
\frac{\partial Q_{jk}}{\partial \phi^{(0)i}}(\phi^{(0)i})
A_1^{(0)i} A_1^{(0)j} A_1^{(0)k}
+  \frac{1}{2} \left(
- \frac{\partial J^k{}_{j}}{\partial \phi^{(0)i}} 
+ \frac{\partial J^k{}_{i}}{\partial \phi^{(0)j}} \right)(\phi^{(0)i})
A_1^{(0)i} A_1^{(0)j} B_{1k}^{(0)}
\nonumber \\ 
&& \quad \quad \ 
+  \frac{1}{2} 
\frac{\partial P^{jk}}{\partial \phi^{(0)i}} (\phi^{(0)i})
A_1^{(0)i} B_{1j}^{(0)} B_{1k}^{(0)}.
\label{nonbv3daction}
\end{eqnarray}
%
%
Since by the dimensional reduction, the fields reduce to
\begin{eqnarray}
&& \phi^{(0)i}(\sigma^1, \sigma^2, \sigma^3) 
= \phizerosub^{(0)i}(\sigma^1, \sigma^2),
\nonumber \\
&& A_1^{(0)i}(\sigma^1, \sigma^2, \sigma^3) 
= \aonesub^{(0)i}(\sigma^1, \sigma^3) 
+ \theta^2 \azerosub^{(0)i}(\sigma^1, \sigma^3),  \nonumber \\
&& B_{1i}^{(0)}(\sigma^1, \sigma^2, \sigma^3) 
= \bonesub_i^{(0)}(\sigma^1, \sigma^3) 
+ \theta^2 \bezerosub_i^{(0)}(\sigma^1, \sigma^3),  \nonumber \\
&& B_{2i}^{(0)}(\sigma^1, \sigma^2, \sigma^3) 
= \btwosub_{i}^{(0)}(\sigma^1, \sigma^3) 
+ \theta^2 \beonesub_{i}^{(0)}(\sigma^1, \sigma^3), 
\end{eqnarray}
the action (\ref{nonbv3daction}) reduces to
\begin{eqnarray}
S_b^{(0)}
%
%
%
%
&=& \int_{S^1} d \sigma^2 \int_{\Pi T \Sprime} 
\frac{1}{2} \left( \beonesub_i^{(0)} \dtwo \phizerosub^{(0)i}
+ \bonesub_i^{(0)} \dtwo \azerosub^{(0)i}
+ \aonesub^{(0)i} \dtwo \bezerosub_i^{(0)} \right) 
- J^i{}_{j} \aonesub^{(0)j} \beonesub_i^{(0)}
+ P{}^{ij} \bonesub_i^{(0)} \beonesub_j^{(0)}
\nonumber \\
&& 
+  \frac{1}{2} \left(
\left(\frac{\partial Q_{jk}}{\partial \phizerosub^{(0)i}} 
+ \frac{\partial Q_{ij}}{\partial \phizerosub^{(0)k}} 
+ \frac{\partial Q_{ki}}{\partial \phizerosub^{(0)j}} 
\right) \azerosub^{(0)k} + 
\left(
- \frac{\partial J^k{}_{j}}{\partial \phizerosub^{(0)i}} 
+ \frac{\partial J^k{}_{i}}{\partial \phizerosub^{(0)j}} \right)
 \bezerosub_k^{(0)} \right)
\aonesub^{(0)i} \aonesub^{(0)j}
\nonumber \\
&& 
+  \left( 
\left(
\frac{\partial J^k{}_{j}}{\partial \phizerosub^{(0)i}} 
- \frac{\partial J^k{}_{i}}{\partial \phizerosub^{(0)j}} \right)
\azerosub^{(0)j} - \frac{\partial P^{jk}}{\partial \phizerosub^{(0)i}} 
\bezerosub_j^{(0)}
\right) 
\aonesub^{(0)i} \bonesub_k^{(0)}
+ \frac{1}{2} \left( \frac{\partial P^{jk}}{\partial \phizerosub^{(0)i}} 
\azerosub^{(0)i}
\right)
\bonesub_j^{(0)} \bonesub_k^{(0)}
\nonumber \\
&& 
- \left(J^i{}_{j} \azerosub^{(0)j} 
+ P{}^{ij}  \bezerosub_j^{(0)} \right) \btwosub_i^{(0)},
%
%
%
%
\label{S0reduction}
\end{eqnarray}
up to total derivative terms.  
Therefore the action $S_{R}^{(0)}$ of a 2D topological field theory 
is 
\begin{eqnarray}
S_R^{(0)} &=& S_0^{(0)} + S_1^{(0)},
\nonumber \\
S_0^{(0)} &=& 
\int_{\Pi T \Sprime} 
\frac{1}{2} \left( \beonesub_i^{(0)} \dtwo \phizerosub^{(0)i}
+ \bonesub_i^{(0)} \dtwo \azerosub^{(0)i}
+ \aonesub^{(0)i} \dtwo \bezerosub_i^{(0)} \right)
\nonumber \\
S_1^{(0)} &=& 
\int_{\Pi T \Sprime} 
- J^i{}_{j} \aonesub^{(0)j} \beonesub_i^{(0)}
+ P{}^{ij} \bonesub_i^{(0)} \beonesub_j^{(0)}
\nonumber \\
&& 
+  \frac{1}{2} \left(
\left(\frac{\partial Q_{jk}}{\partial \phizerosub^{(0)i}} 
+ \frac{\partial Q_{ij}}{\partial \phizerosub^{(0)k}} 
+ \frac{\partial Q_{ki}}{\partial \phizerosub^{(0)j}} 
\right) \azerosub^{(0)k} + 
\left(
- \frac{\partial J^k{}_{j}}{\partial \phizerosub^{(0)i}} 
+ \frac{\partial J^k{}_{i}}{\partial \phizerosub^{(0)j}} \right)
 \bezerosub_k^{(0)} \right)
\aonesub^{(0)i} \aonesub^{(0)j}
\nonumber \\
&& 
+  \left( 
\left(
\frac{\partial J^k{}_{j}}{\partial \phizerosub^{(0)i}} 
- \frac{\partial J^k{}_{i}}{\partial \phizerosub^{(0)j}} \right)
\azerosub^{(0)j} - \frac{\partial P^{jk}}{\partial \phizerosub^{(0)i}} 
\bezerosub_j^{(0)}
\right) 
\aonesub^{(0)i} \bonesub_k^{(0)}
+ \frac{1}{2} \left( \frac{\partial P^{jk}}{\partial \phizerosub^{(0)i}} 
\azerosub^{(0)i}
\right)
\bonesub_j^{(0)} \bonesub_k^{(0)}
\nonumber \\
&& 
- \left(J^i{}_{j} \azerosub^{(0)j} 
+ P{}^{ij}  \bezerosub_j^{(0)} \right) \btwosub_i^{(0)}.
\end{eqnarray}

Next we formulate the action $S_{R}$ by the AKSZ formulation.
%
We define $S_R = S_0 + S_1$
where $S_0$ and $S_1$ are AKSZ actions for $S_0^{(0)}$ and $S_1^{(0)}$, respectively.  
$S_0$ is easily derived after substituting (\ref{fielddimred}) to 
(\ref{boundarysigma}); 
\begin{eqnarray}
S_0 
&=& 
\int_{\Pi T \Sprime} 
\frac{1}{2} \left( \beone_i \dtwo \phizero^i
- \btwo_i d \phiminus^i 
+ \bone_i \dtwo \azero^i
+ \aone^i \dtwo \bezero_i \right)
\label{S0AKSZ}
\end{eqnarray}
up to total derivative terms.  The condition $\sbv{S_1}{S_1}=0$ comes from (\ref{S0AKSZ}) and $\sbv{S_R}{S_R}=0$.  
We introduce an {\it negative total degree}, 
which is defined as one for $\phiminus$, 
and zero for the other fields.  
We can expand $S_1$ for the negative total degree such as 
$S_1 = \sum_{p=0}^{\infty} S_1^{[p]}$, where 
\begin{equation}
S_1^{[p]} = \int_{\Pi T \Sprime}
\phiminus^{i_1} \cdots \phiminus^{i_p}
\calL_{i_1 \cdots i_p}^{[p]}(\phizero, \aone^i, \azero, \bone_i, 
\bezero_i, \btwo_{i}, \beone_{i})
\end{equation}
are the negative total degree $p$ terms.  
Therefore
\begin{eqnarray}
S_R = S_0 + \sum_{p=0}^{\infty} S_1^{[p]}.
\label{bfreduct}
\end{eqnarray}
Here we write the first two actions $S_1^{[0]}$ and $S_1^{[1]}$ 
with the negative total degree zero and one by substituting (\ref{fielddimred}) to (\ref{boundarysigma}),
\begin{eqnarray}
S_1^{[0]} 
&=& 
\int_{\Pi T \Sprime} 
- J^i{}_{j} \aone^j \beone_i
+ P{}^{ij} \bone_i \beone_j
\nonumber \\
&& 
+  \frac{1}{2} \left(
\left(\frac{\partial Q_{jk}}{\partial \phizero^{i}} 
+ \frac{\partial Q_{ij}}{\partial \phizero^{k}} 
+ \frac{\partial Q_{ki}}{\partial \phizero^{j}} 
\right) \azero^k + 
\left(
- \frac{\partial J^k{}_{j}}{\partial \phizero^i} 
+ \frac{\partial J^k{}_{i}}{\partial \phizero^j} \right)
 \bezero_k \right)
\aone^i \aone^j 
\nonumber \\
&& 
+  \left( 
\left(
\frac{\partial J^k{}_{j}}{\partial \phizero^i} 
- \frac{\partial J^k{}_{i}}{\partial \phizero^j} \right)
\azero^j - \frac{\partial P^{jk}}{\partial \phizero^i} \bezero_j
\right) 
\aone^i \bone_k 
+ \frac{1}{2} \left( \frac{\partial P^{jk}}{\partial \phizero^i} \azero^i
\right)
\bone_j \bone_k
\nonumber \\
&& 
- \left(J^i{}_{j} \azero^j + P{}^{ij}  \bezero_j \right) \btwo_i ,
\\
%
S_1^{[1]}
&=& 
\int_{\Pi T \Sprime} 
\phiminus^l 
\Biggr[ \frac{\partial J^i{}_{j}}{\partial \phizero^l}
\btwo_i \aone^j 
+ \frac{\partial P{}^{ij}}{\partial \phizero^l}
\btwo_i \bone_j
- \frac{1}{2} 
\frac{\partial^2 Q_{jk}}{\partial \phizero^{i} \partial \phizero^l} 
\aone^i \aone^j \aone^k 
\nonumber \\
&& 
- \frac{1}{2} 
\frac{\partial}{\partial \phizero^l} \left(
- \frac{\partial J^k{}_{j}}{\partial \phizero^i} 
+ \frac{\partial J^k{}_{i}}{\partial \phizero^j} \right)
\aone^i \aone^j \bone_k 
- \frac{1}{2} \frac{\partial^2 P^{jk}}{\partial \phizero^i 
\partial \phizero^l} 
\aone^i \bone_j \bone_k 
\Biggl].
\end{eqnarray}
$S_1^{[p]}$ for $p > 1$ are recursively derived from the master equation
$\sbv{S_1}{S_1}= \sum_{p=0}^{\infty} \{\sbv{S_1}{S_1}\}^{[p]} = 0$.  
It should be noticed that since a target space $M$ has finite dimensions,
$S_1^{[p]}$ is nonzero for only a finite number of $p$.  
This action is a special case of 
a nonlinear gauge theory with $2$-forms
(a generalization of the Poisson sigma model)
analyzed in the paper 
\cite{Ikeda:2004gp}\cite{Batalin:2001fh}\cite{Batalin:2001fc}.

\subsection{$H \neq 0$}
Here we consider $H \neq 0$ case.  A 2D topological field theory of 
twisted generalized complex geometry is derived in a similar way in subsection 4.1 from $H$-terms in section 3.2: 
\begin{eqnarray}
S_R &=& S_0 + \sum_{p=0}^{\infty} S_1^{[p]} ;
\\
S_1^{[0]} 
&=& 
\int_{\Pi T \Sprime} 
- J^i{}_{j} \aone^j \beone_i
+ P{}^{ij} \bone_i \beone_j
\nonumber \\
&& 
+  \frac{1}{2} \left(
\left(3 H_{ijk} + \frac{\partial Q_{jk}}{\partial \phizero^{i}} 
+ \frac{\partial Q_{ij}}{\partial \phizero^{k}} 
+ \frac{\partial Q_{ki}}{\partial \phizero^{j}} 
\right) \azero^k + 
\left(
- \frac{\partial J^k{}_{j}}{\partial \phizero^i} 
+ \frac{\partial J^k{}_{i}}{\partial \phizero^j} \right)
 \bezero_k \right)
\aone^i \aone^j 
\nonumber \\
&& 
+  \left( 
\left(
\frac{\partial J^k{}_{j}}{\partial \phizero^i} 
- \frac{\partial J^k{}_{i}}{\partial \phizero^j} \right)
\azero^j - \frac{\partial P^{jk}}{\partial \phizero^i} \bezero_j
\right) 
\aone^i \bone_k 
+ \frac{1}{2} \left( \frac{\partial P^{jk}}{\partial \phizero^i} \azero^i
\right)
\bone_j \bone_k
\nonumber \\
&& 
- \left(J^i{}_{j} \azero^j + P{}^{ij}  \bezero_j \right) \btwo_i ,
\\
%
S_1^{[1]}
&=& 
\int_{\Pi T \Sprime} 
\phiminus^l 
\Biggr[ \frac{\partial J^i{}_{j}}{\partial \phizero^l}
\btwo_i \aone^j 
+ \frac{\partial P{}^{ij}}{\partial \phizero^l}
\btwo_i \bone_j
- \frac{1}{2} 
\frac{\partial}{\partial \phizero^l}
\left(H_{ijk} +
\frac{\partial Q_{jk}}{\partial \phizero^{i}} \right)
\aone^i \aone^j \aone^k 
\nonumber \\
&& 
- \frac{1}{2} 
\frac{\partial}{\partial \phizero^l} \left(
- \frac{\partial J^k{}_{j}}{\partial \phizero^i} 
+ \frac{\partial J^k{}_{i}}{\partial \phizero^j} \right)
\aone^i \aone^j \bone_k 
- \frac{1}{2} \frac{\partial^2 P^{jk}}{\partial \phizero^i 
\partial \phizero^l} 
\aone^i \bone_j \bone_k 
\Biggl],
\end{eqnarray}
and $S_1^{[p]}$ for $p > 1$ are recursively derived from $(S_R, S_R)$.

Also, from $b$-invariant $H$-terms in section 3.3,
\begin{eqnarray}
S_R &=& S_0 + \sum_{p=0}^{\infty} S_1^{[p]} ;
\\
S_1^{[0]} 
&=& 
\int_{\Pi T \Sprime} 
- J^i{}_{j} \aone^j \beone_i
+ P{}^{ij} \bone_i \beone_j
\nonumber \\
&& 
+  \frac{1}{2} \left(
\left(3 J^l{}_i H_{jkl} + \frac{\partial Q_{jk}}{\partial \phizero^{i}} 
+ \frac{\partial Q_{ij}}{\partial \phizero^{k}} 
+ \frac{\partial Q_{ki}}{\partial \phizero^{j}} 
\right) \azero^k + 
\left(- P^{kl} H_{jkl}
- \frac{\partial J^k{}_{j}}{\partial \phizero^i} 
+ \frac{\partial J^k{}_{i}}{\partial \phizero^j} \right)
 \bezero_k \right)
\aone^i \aone^j 
\nonumber \\
&& 
+  \left( 
\left(P^{kl} H_{jkl} +
\frac{\partial J^k{}_{j}}{\partial \phizero^i} 
- \frac{\partial J^k{}_{i}}{\partial \phizero^j} \right)
\azero^j - \frac{\partial P^{jk}}{\partial \phizero^i} \bezero_j
\right) 
\aone^i \bone_k 
+ \frac{1}{2} \left( \frac{\partial P^{jk}}{\partial \phizero^i} \azero^i
\right)
\bone_j \bone_k
\nonumber \\
&& 
- \left(J^i{}_{j} \azero^j + P{}^{ij}  \bezero_j \right) \btwo_i ,
\\
%
S_1^{[1]}
&=& 
\int_{\Pi T \Sprime} 
\phiminus^l 
\Biggr[ \frac{\partial J^i{}_{j}}{\partial \phizero^l}
\btwo_i \aone^j 
+ \frac{\partial P{}^{ij}}{\partial \phizero^l}
\btwo_i \bone_j
- \frac{1}{2} 
\frac{\partial}{\partial \phizero^l}
\left(J^m{}_i H_{jkm} +
\frac{\partial Q_{jk}}{\partial \phizero^{i}} \right)
\aone^i \aone^j \aone^k 
\nonumber \\
&& 
- \frac{1}{2} 
\frac{\partial}{\partial \phizero^l} \left(
- P^{km} H_{jkm}
- \frac{\partial J^k{}_{j}}{\partial \phizero^i} 
+ \frac{\partial J^k{}_{i}}{\partial \phizero^j} \right)
\aone^i \aone^j \bone_k 
- \frac{1}{2} \frac{\partial^2 P^{jk}}{\partial \phizero^i 
\partial \phizero^l} 
\aone^i \bone_j \bone_k 
\Biggl],
\end{eqnarray}
and $S_1^{[p]}$ for $p > 1$ are recursively derived from $(S_R, S_R)$.  
\section{Two Special Reductions to Complex Geometry and Symplectic Geometry}
\noindent
In this section, we consider two special reductions related to complex geometry and of symplectic geometry.  

\subsection{Complex geometry}
\noindent
First we consider our model in complex geometry, which is the case that 
$P=Q=H=0$ in the action (\ref{bfreduct}).  
We redefine superfields as
\begin{eqnarray}
&& \phizero^i = \phizero^i, \qquad 
\phiminus^{i} = \lambda \phiminusp^{i},  
\nonumber \\
&& \aone^i = \frac{1}{2} \aonep^i,
\qquad \azero^i = \lambda \azerop^i,  
\nonumber \\
&& \bone_i = \lambda \bonep_i,
\qquad \bezero_i = - \bezerop_i,  
\nonumber \\
&& \btwo_{i} = \lambda \btwop_{i},
\qquad \beone_{i} = \frac{1}{2} \beonep_{i},
\end{eqnarray}
where $\lambda$ is a constant.  After this redefinition, the action (\ref{bfreduct}) is 
\begin{eqnarray}
S_R &=& S_0 + \sum_{p=0}^{\infty} S_1^{[p]}, 
\label{blikemodel} \\
S_0 
&=& 
\int_{\Pi T \Sprime} 
\frac{1}{4} \left( \beonep_i \dtwo \phizero^i
- \aonep^{i} \dtwo \bezerop_i \right)
+ \frac{\lambda^2}{2} \left(- \btwop_i d \phiminusp^{i}
+ \bonep_i \dtwo \azerop^{i}
\right), \label{kineticterm}
\\
%
S_1^{[0]} 
&=& 
\int_{\Pi T \Sprime} 
\frac{1}{4} \left( J^i{}_{j} \beonep_i \aonep^{j}
+ \frac{\partial J^k{}_{j}}{\partial \phizero^i} 
\bezerop_k \aonep^i \aonep^j \right)
\nonumber \\
&&
~~~ + \lambda^2 \left(\frac{1}{2}
\frac{\partial J^k{}_{j}}{\partial \phizero^i} 
\azerop^j \aonep^i \bonep_k 
- J^i{}_{j} \azerop^j \btwop_i \right), \label{twoterms} 
\\
%
S_1^{[1]}
&=& 
\int_{\Pi T \Sprime} 
\lambda \phiminus^l 
\Biggr[ \frac{\lambda}{2} 
\frac{\partial J^i{}_{j}}{\partial \phizero^l} \btwop_i \aonep^j 
+ \frac{\lambda}{4} 
\frac{\partial^2 J^k{}_{j}}{\partial \phizero^l \partial \phizero^i} 
\aonep^i \aonep^j \bonep_k 
\Biggl], 
\end{eqnarray}
and $S_1^{[p]}$ has at least the higher order of $\lambda$ than $\lambda^p$ because $\phiminus^{i} = \lambda \phiminusp^{i}$.  
We can take the limit $\lambda \longrightarrow 0$ with preserving the complex structure.  
$S_1^{[p]}$ for $p > 0$ reduces to zero, 
and 
%
%
%
the 2D action is 
\begin{eqnarray}
S_{RJ}
&=& \frac{1}{4}  \int_{\Pi T \Sprime} 
\beone_i \dtwo \phizero^i
- \aone^i \dtwo \bezero_i
+ J^i{}_{j} \beone_i \aone^j 
+ \frac{\partial J^i{}_{k}}{\partial \phizero^j} \bezero_i
\aone^j \aone^k.
\label{breduction}
\end{eqnarray}
This action is nothing but the $B$ model action (\ref{bmodel})
up to a total derivative and the all over factor $\frac{1}{4}$, 
which depends on only $J^i{}_{j}$.  
The master equation $(S_{bJ}, S_{bJ}) = 0$ 
impose the condition that $J^i{}_{j}$
is a complex structure.  

We make a comment about the difference between the action (\ref{blikemodel}) with a finite $\lambda$ and the B model action (\ref{breduction}) with $\lambda \rightarrow 0$.  
Following the well-known method in \cite{Alexandrov:1995kv}, we can see that the topological string theory has to be deformed by the other terms in (\ref{blikemodel}) than in the B model.   
In the calculation of \cite{Alexandrov:1995kv}, we may locally take the complex structure as a constant, and the kinetic terms (\ref{kineticterm}) and two terms in (\ref{twoterms}) without the derivatives of $J^i{}_{j}$ are only different parts from those in the B model.  
Here it should be noted that although these deformed parts may seem to decouple to the B model part, the interactions between them can come from the non-constant metric.  
These deformed parts couple to only the metric on the bosonic space of $\phizero^i$, which is independent of $\phiminus^{i}$, 
because there is no metric with fermionic indices on the fermionic space of $\phiminus^{i}$.  
So these deformed parts can be seen as a topological theory with only B field-like couplings on the fermionic space of $\phiminus^{i}$.  
Physically, we may assume that there is no topological information along fermionic directions, although this situation with no metric is special.   
Therefore in this assumption, we can see that our action (\ref{blikemodel}) is equivalent to topological string theory, called topological B model.   
As a future work, it would be interesting to check this equivalence more carefully.

\subsection{Symplectic geometry}
Next we consider our model in symplectic geometry, which is the case that 
$J=H=0$ in the action (\ref{bfreduct}).  
We redefine superfields as
\begin{eqnarray}
&& \phizero^i = \phizero^i, \qquad 
\phiminus^{i} = \mu \phiminusp^{i},  
\nonumber \\
&& \aone^i = \mu \aonep^i,
\qquad 
\azero^i = \azerop^i,  
\nonumber \\
&& \bone_i = \frac{1}{2} \bonep_i,
\qquad 
\bezero_i = - \mu \bezerop_i,  
\nonumber \\
&& \btwo_{i} = \mu \btwop_{i},
\qquad 
\beone_{i} = \frac{1}{2} \beonep_{i},
\end{eqnarray}
where $\mu$ is a constant.  
After this redefinition, the 2D action (\ref{bfreduct}) reduces to 
\begin{eqnarray}
%
S_R &=& S_0 + \sum_{p=0}^{\infty} S_1^{[p]},
\nonumber \\
S_0 
&=& 
\int_{\Pi T \Sprime} 
\frac{1}{4} \left( \beonep_i \dtwo \phizero^i
+ \bonep_i \dtwo \azerop^i \right)
+ \frac{\mu^2}{2} \left(- \btwop_i d \phiminusp^i 
- \aonep^i \dtwo \bezerop_i \right)
\\
S_1^{[0]} 
&=& 
\int_{\Pi T \Sprime} 
\frac{1}{4} \left(
P{}^{ij} \bonep_i \beonep_j
+ \frac{1}{2} \frac{\partial P^{jk}}{\partial \phizero^i} \azerop^i
\bonep_j \bonep_k \right)
\\
&&
+ \mu^2 \left( \frac{1}{2} 
\left(\frac{\partial Q_{jk}}{\partial \phizero^{i}} 
+ \frac{\partial Q_{ij}}{\partial \phizero^{k}} 
+ \frac{\partial Q_{ki}}{\partial \phizero^{j}} 
\right) \azerop^k \aonep^i \aonep^j
- \frac{1}{2} \frac{\partial P^{jk}}{\partial \phizero^i} \bezerop_j
\aonep^i \bonep_k 
+ P{}^{ij}  \bezerop_j \btwop_i \right),
\nonumber \\
%
S_1^{[1]}
&=& 
\int_{\Pi T \Sprime} 
\mu \phiminus^l 
\Biggr[ \frac{\mu}{2} 
\frac{\partial P{}^{ij}}{\partial \phizero^l}
\btwo_i \bone_j
- \frac{\mu^3}{2} 
\frac{\partial^2 Q_{jk}}{\partial \phizero^{i} \partial \phizero^l} 
\aonep^i \aonep^j \aonep^k 
- \frac{\mu}{8} \frac{\partial^2 P^{jk}}{\partial \phizero^i 
\partial \phizero^l} 
\aonep^i \bonep_j \bonep_k 
\Biggl],
\end{eqnarray}
and $S_1^{[p]}$ is at least the higher order of $\lambda$ than $\lambda^p$ because $\phiminus^{i} = \mu \phiminusp^{i}$.  

After taking the limit $\mu \longrightarrow 0$ with preserving the symplectic structure, 
$S_1^{[p]}$ for $p > 0$ reduces to zero,
and the 2D action is
\begin{eqnarray}
S_{RP}
&=& \frac{1}{4}  \int_{\Pi T \Sprime} 
\beone_i \dtwo \phizero^i
+ \bone_i \dtwo \azero^i
+ P{}^{ij} \bone_i \beone_j
+ \frac{1}{2} \frac{\partial P^{jk}}{\partial \phizero^i} \azero^i
\bone_j \bone_k.
\label{areduction}
\end{eqnarray}
The BV condition $\sbv{S_{bP}}{S_{bP}} = 0$ is satisfied if and only if 
$P^{ij}$ is a Poisson structure (the inverse of a symplectic structure)
(\ref{Poissoncondition}).  
It should be noticed that although this action (\ref{areduction}) 
depends on only a symplectic structure
$P^{ij}$, this action is a different realization of the Poisson 
structure from the $A$ model (\ref{apmodel}), because we can also check that this model is not equivalent to topological string theory following the similar way as in \cite{Alexandrov:1995kv}.   

\section{Conclusions and Discussion}
\noindent
We have constructed a topological field theory 
with a generalized complex structure in three dimensions and two dimensions
using the AKSZ formulation.  
Our model reduces to B model in a limit if the 
generalized complex structure is only a complex structure, 
although Zucchini model reduces to A model in the limit that 
the generalized complex structure is only a symplectic structure.

It would be interesting to check that the Zucchini model and 
our model are equivalent to 
a topological string theory with a generalized complex structure
\cite{Chuang:2006vt}\cite{Zucchini:2006ii}, which is constructed from 
the twisted $N=(2,2)$ supersymmetric sigma model with 
a non-trivial $B$ field.

%



\section*{Appendix A. Generalized Complex Structure}
\noindent
In this appendix A, we summarize a generalized complex structure, 
based on description of section 3 in \cite{LMTZ} 
and section 2 in \cite{Zucchini:2004ta}.

Let $M$ be a manifold of even dimension $d$ with 
a local coordinate $\{ \phi^i \}$. We consider the vector bundle 
$TM \oplus T^*M$. 
We denote a section as $X + \xi \in C^\infty(TM\oplus T^*M)$ 
where $X \in C^\infty(TM)$ and $\xi \in C^\infty(T^*M)$.

$TM \oplus T^* M$ is equipped with a natural indefinite metric of signature  
$(d,d)$ defined by 
\begin{eqnarray}
\langle X + \xi , Y+\eta \rangle= \frac{1}{2} (i_X \eta + i_Y \xi),
\end{eqnarray}
for $X+\xi, Y+\eta \in C^\infty (TM \oplus T^* M)$, where $i_V$ 
is an interior product with a vector field $V$. 
In the Cartesian coordinate $(\partial/\partial \phi^i, d \phi^i)$, 
The metric is written as follows:
\begin{eqnarray}
{\calI} = \left(\matrix{0 & 1_d &\cr
                          1_d & 0 &\cr}\!\!\!\!\!\!\!\right),
\end{eqnarray}

We define a {\it Courant bracket} on $TM \oplus T^* M$ as follows:
\begin{eqnarray}
[ X + \xi, Y + \eta] = [X,Y] + \calL_X \eta - \calL_Y \xi 
- \frac{1}{2} d_M (i_X\eta-i_Y\xi),
\end{eqnarray}
with $X+\xi, Y+\eta\in C^\infty(TM\oplus T^* M)$, where $\calL_V$ 
denotes Lie derivation with respect a vector field $V$ and $d_M$ is 
the exterior differential of $M$.
This bracket is antisymmetric but do not satisfy the Jacobi identity.
We may consider a so called Dorfman bracket as follows:
\begin{eqnarray}
(X+\xi) \circ (Y+\eta) = [X,Y]+ \calL_X \eta - i_Y d \xi,
\end{eqnarray}
which satisfies the Jacobi identity but is not antisymmetric.
Antisymmetrization of a Dorfman bracket coincides with a Courant bracket.


A {\it generalized almost complex structure} $\calJ$ is a section of 
$C^\infty ({\rm End}(TM\oplus T^* M))$, which is an isometry of 
the metric $\langle\,,\rangle$, 
$
{\calJ}^* \calI \calJ = \calI,
$
and satisfies 
\begin{eqnarray}
{\calJ}^2 = -1.
\end{eqnarray}

A $b$-transformation is an isometry defined by 
\begin{eqnarray}
\exp(b)( X + \xi)= X + \xi + i_X b,
\end{eqnarray}
where $b \in C^\infty(\wedge^2 T^* M)$ is a $2$--form. 
A Courant bracket is covariant under the $b$-transformation
\begin{eqnarray}
[\exp(b) (X + \xi), \exp(b)(Y + \eta)] = \exp(b)[X + \xi, Y + \eta],
\end{eqnarray}
if the $2$--form $b$ is closed.
%
The $b$-transform of $\calJ$ is defined by 
\begin{eqnarray}
\hat{\calJ}=\exp(-b) {\calJ} \exp(b).
\end{eqnarray}

$\calJ$ has the $\pm \sqrt{{-1}}$ eigenbundles
because ${\calJ}^2 = -1$, 
In order to divide $TM \oplus T^* M$ to each eigenbundle,
we need complexification of $TM \oplus T^* M$, 
$(TM \oplus T^* M) \otimes \C$.
The projectors on the eigenbundles are defined by 
\begin{eqnarray}
\Pi_{\pm} = \hbox{$1\over 2$}(1\mp\sqrt{{-1}}{\calJ}).
\end{eqnarray}
The generalized almost complex structure $\calJ$ is integrable if 
\begin{eqnarray}
\Pi_{\mp}[\Pi_{\pm}(X+\xi),\Pi_{\pm}(Y+\eta)]=0,
\label{integrability}
\end{eqnarray}
for any $X+\xi, Y+\eta \in C^\infty(TM \oplus T^* M)$,
where the bracket is the Courant bracket.
Then $\calJ$ is called a {\it generalized complex structure}. 
Integrability is equivalent to the single statement
\begin{eqnarray}
N(X+\xi,Y+\eta)=0,
\end{eqnarray}
for all $X+\xi, Y+\eta\in C^\infty(TM\oplus T^* M)$, 
where $N$ is the generalized Nijenhuis tensor defined by 
\begin{eqnarray}
N(X+\xi,Y+\eta)
&=& [X+\xi,Y+\eta] -[{\calJ}(X+\xi), {\calJ}(Y+\eta)]
+{\calJ}[{\calJ}(X+\xi),Y+\eta]
\nonumber \\
&&\, +{\calJ}[X+\xi, {\calJ}(Y+\eta)].
\end{eqnarray}
The $b$-transform $\hat{\calJ}$ of a generalized complex structure $\calJ$ is 
a generalized complex structure if the $2$--form $b$ is closed. 

We decompose a generalized almost complex structure 
$\calJ$ in coordinate form as follows 
\begin{eqnarray}
{\calJ} = \left(\matrix{J& P&\cr
                          Q& K &\cr}\!\!\!\!\!\!\!\right),
\label{gcstcomp}
\end{eqnarray}
where $J, K \in C^\infty(TM \otimes T^* M)$, $P\in C^\infty(\wedge^2 TM)$, 
$Q\in C^\infty(\wedge^2 T^* M)$.

Then the conditions 
$
{\calJ}^* \calI \calJ = \calI,
$
and 
${\calJ}^2 = -1$ derive 
\begin{eqnarray}
&&
K_j{}^i = -J^{i}{}_j
\nonumber \\
&& 
J^i{}_k J^k{}_j + P^{ik}Q_{kj} + \delta^i{}_j = 0,
\nonumber \\
&& 
J^i{}_k P^{kj} + J^j{}_k P^{ki} = 0,
\nonumber \\
&& 
Q_{ik} J^k{}_j + Q_{jk} J^k{}_i = 0,
\label{J2zero}
\end{eqnarray}
where 
\begin{eqnarray}
&& 
P^{ij}+P^{ji}=0,
\nonumber \\
&& 
Q_{ij}+Q_{ji}=0.
\label{PQantisym}
\end{eqnarray}
The integrability condition (\ref{integrability}) is equivalent to the 
following condition 
\begin{eqnarray}
\calA^{ijk}= \calB_i{}^{jk} = \calC_{ij}{}^k = \calD_{ijk} = 0,
\label{intcoord}
\end{eqnarray}
where
\begin{eqnarray}
\calA^{ijk} &=& P^{il} \partial_l P^{jk} 
+ P^{jl} \partial_l P^{ki} + P^{kl} \partial_l P^{ij},
\nonumber \\
\calB_i{}^{jk} &=& J^l{}_i \partial_l P^{jk}
+ P^{jl} (\partial_i J^k{}_l - \partial_l J^k{}_i)
+ P^{kl} \partial_l J^j{}_i - J^j{}_l \partial_i P^{lk},
\nonumber \\
\calC_{ij}{}^k &=& 
J^l{}_i \partial_l J^k{}_j - J^l{}_j \partial_l J^k{}_i
- J^k{}_l \partial_i J^l{}_j + J^k{}_l \partial_j J^l{}_i
\nonumber \\
&&
+ P^{kl} (\partial_l Q_{ij} + \partial_i Q_{jl} + \partial_j Q_{li}),
\nonumber \\
\calD_{ijk} &=& J^l{}_{i} (\partial_l Q_{jk} + \partial_k Q_{lj})
+ J^l{}_j (\partial_l Q_{ki} + \partial_i Q_{lk})
\nonumber \\
&&
+ J^l{}_k (\partial_l Q_{ij} + \partial_j Q_{li})
- Q_{jl} \partial_i J^l{}_k - Q_{kl} \partial_j J^l{}_i 
- Q_{il} \partial_k J^l{}_j.
\end{eqnarray}
Here $\partial_i$ is a differentiation with respect to $\phi^i$.
The $b$--transform is 
\begin{eqnarray}
\hat J^i{}_j &=& J^i{}_j - P^{ik}b_{kj},
\nonumber \\
\hat P^{ij} &=& P^{ij},
\nonumber \\
\hat Q_{ij} &=& Q_{ij} + b_{ik} J^k{}_j - b_{jk} J^k{}_i 
+ P^{kl} b_{ki} b_{lj}.
\label{btrans}
\end{eqnarray}
where $b_{ij} + b_{ji} =0$.

The usual complex structures $J$ is embedded in 
generalized complex structures 
as the special form 
\begin{eqnarray}
 {\calJ} = \left (\matrix{J & 0 &\cr
                           0 & -{}^t \! J&\cr}\!\!\!\!\!\!\right).
\end{eqnarray}
Indeed, one can check this form satisfies conditions,
(\ref{J2zero}) and (\ref{intcoord}) 
if and only if $J$ is a complex structure.
Similarly, the usual symplectic structures $Q$ is obtained as 
the special form of generalized complex structures 
\begin{eqnarray}
{\calJ} = \left (\matrix{0 & -Q^{-1} &\cr
                          Q & 0&\cr}\!\!\!\!\!\!\right).
\end{eqnarray}
This satisfies (\ref{J2zero}) and (\ref{intcoord}) 
if and only if $Q$ is a symplectic structure, i.~e.~it is closed. 
Other exotic examples exist. 
There exists manifolds which cannot support any complex or 
symplectic structure, 
but admit generalized complex structures. 

The Courant bracket on $TM \oplus T^*M$ can be modified by a closed 
$3$--form.
Let $H \in C^\infty(\wedge^3 T^*M)$ be a closed $3$--form.
We define the $H$ twisted Courant brackets by
\begin{eqnarray}
[X + \xi, Y + \eta]_H = [X + \xi , Y + \eta] + i_X i_Y H,
\end{eqnarray}
where $X+\xi, Y+\eta \in C^\infty(TM\oplus T^* M)$.
Under the $b$-transform with $b$ a closed $2$--form, 
\begin{eqnarray}
[\exp(b) (X + \xi), \exp(b)(Y + \eta)] = \exp(b)[X + \xi, Y + \eta],
\end{eqnarray}
holds with the brackets $[\,,]$ replaced by $[\,,]_H$. 
For a non closed $b$, one has
\begin{eqnarray}
[\exp(b)(X+\xi),\exp(b)(Y+\eta)]_{H - d_Mb}=\exp(b)[X+\xi,Y+\eta]_H.
\end{eqnarray}
So, the $b$-transformation shifts $H$ by the exact $3$--form $d_M b$:
\begin{eqnarray}
\hat H=H - d_M b.
\label{btransj}
\end{eqnarray}
%
%
One can define an $H$ twisted generalized Nijenhuis tensor $N_H$ as follows
\begin{eqnarray}
N(X+\xi,Y+\eta)
&=& [X+\xi,Y+\eta]_H -[{\calJ}(X+\xi), {\calJ}(Y+\eta)]_H
+{\calJ}[{\calJ}(X+\xi),Y+\eta]_H
\nonumber \\
&&\, +{\calJ}[X+\xi, {\calJ}(Y+\eta)]_H,
\end{eqnarray}
by using the brackets $[\,,]_H$ instead of $[\,,]$.
A generalized almost complex structure $\cal J$ is $H$ integrable if
\begin{eqnarray}
N_H(X+\xi,Y+\eta)=0,
\end{eqnarray}
for all $X+\xi,Y+\eta \in C^\infty(TM\oplus T^* M)$.
Then we call $\cal J$ an {\it twisted generalized complex structure}. 

The $H$ integrability conditions is as follows:
\begin{eqnarray}
\calA_H{}^{ijk} = \calB_H{}_i{}^{jk} = \calC_H{}_{ij}{}^k =
\calD_H{}_{ijk} = 0,
\end{eqnarray}
where
\begin{eqnarray}
&& 
\calA_H{}^{ijk} = \calA^{ijk},
\nonumber \\
&& 
\calB_H{}_i{}^{jk}
= \calB_i{}^{jk} + P^{jl}P^{km}H_{ilm}
\nonumber \\
&& 
\calC_H{}_{ij}{}^k 
= \calC_{ij}{}^k - J^l{}_i P^{km} H_{jlm} + J^l{}_j P^{km} H_{ilm},
\nonumber \\
&& 
\calD_H{}_{ijk} = \calD_{ijk} - H_{ijk} 
+ J^l{}_i J^m{}_j H_{klm} + J^l{}_j J^m{}_k H_{ilm}
+ J^l{}_k J^m{}_i H_{jlm}.
\label{tgcstcomp}
\end{eqnarray}

\section*{Appendix B. AKSZ Formulation of Batalin-Vilkovisky Formalism}
\noindent
In the appendix B, we review the AKSZ formulation
in any dimension \cite{Ikeda:2006wd}.
In order to construct and analyze topological field theories 
systematically, 
it is useful to use Batalin-Vilkovisky formalism.
The geometric structure of the AKSZ formulation 
is called Batalin-Vilkovisky Structures.

\subsection*{B-1. Batalin-Vilkovisky Structures on Graded Vector Bundles}
\noindent
%
%
Let $M$ be a smooth manifold in $d$ dimensions.
If we consider 
We define a {\it supermanifold} $\Pi T^* M$.
Mathematically, $\Pi T^* M$, whose bosonic part is $M$, is defined as 
a cotangent bundle with reversed parity of the fiber.
That is, a base manifold $M$ has a Grassman even coordinate and 
the fiber of $\Pi T^* M$ has a Grassman odd coordinate.
We introduce a grading called {\it total degrees}, which is denoted $|F|$
for a function $F$.
The coordinates of the base manifold have grade zero and 
the coordinates of the fiber have grade one.
Similarly, we can define $\Pi T M$ for a tangent bundle $T M$.
$\Pi T M$ is also called a supermanifold.

We must consider more general assignments 
for the degree of the fibers of $T^* M$ or $T M$.
For an 
integer $p$, we define $T^* [p] M$, which is called 
a {\it graded cotangent bundle}.
$T^* [p] M$ is a cotangent bundle, whose fiber has the degree $p$.
This degree is also called the {\it total degree}.
A coordinate of the bass manifold have the total degree zero and 
a coordinate of the fiber have the total degree $p$.
If $p$ is odd, the fiber is Grassman odd, and if $p$ is even, 
the fiber is Grassman even. 
We define a graded tangent bundle $T [p] M$ in the same way.
%

We consider a vector bundle $E$.
A {\it graded vector bundle} $E[p]$ is defined in the similar way.
$E[p]$ is a vector bundle whose fiber has a shifted degree by $p$.  
Note that only the degree of fiber is shifted, and 
the degree of base space is not shifted. 

We consider a Poisson manifold $N$ with a Poisson bracket $\{*,*\}$.
If we shift the total degree, 
we can construct a graded manifold (a graded cotangent bundle or
a graded vector bundle) $\tilde{N}$ from $N$.  
Then a Poisson structure $\{*,*\}$ shifts to
a graded Poisson structure by grading of $\tilde{N}$.
The graded Poisson bracket is called an {\it antibracket} 
and denoted by $\sbv{*}{*}$.
$\sbv{*}{*}$ is graded symmetric and satisfies the graded Leibniz
rule and the graded Jacobi identity with respect to grading of the manifold. 
%
The antibracket $\sbv{*}{*}$ with 
 the total degree $- n + 1$ satisfies the following identities:
\begin{eqnarray}
&& \sbv{F}{G} = -(-1)^{(|F| + 1 - n)(|G| + 1 - n)} \sbv{G}{F},
\nonumber \\
&& \sbv{F}{G  H} = \sbv{F}{G} H
+ (-1)^{(|F| + 1 - n)|G|} G \sbv{F}{H},
\nonumber \\
&& \sbv{F  G}{H} = F \sbv{G}{H}
+ (-1)^{|G|(|H| + 1 - n)} \sbv{F}{H} G,
\nonumber \\
&& (-1)^{(|F| + 1 - n)(|H| + 1 - n)} \sbv{F}{\sbv{G}{H}}
+ {\rm cyclic \ permutations} = 0,
\label{BVidentity}
\end{eqnarray}
where $F, G$ and $H$ are functions on $\tilde{N}$, and
$|F|, |G|$ and $|H|$ are total degrees of the functions, respectively.
The graded Poisson structure is also called {\it P-structure}.
If $n=1$, the antibracket is equivalent to the Schouten bracket.
For higher $n$, the antibracket is equivalent to
the Loday bracket \cite{Loday} with the degree $-n+1$.

Typical examples of Poisson manifold $N$ are a cotangent bundle $T^* M$ 
and a vector bundle $E \oplus E^*$.  
%
First we consider a cotangent bundle $T^* M$.
Since $T^* M$ has a natural symplectic structure,
we can define a Poisson bracket induced from the 
symplectic structure.
If we take a local coordinate $\phi^i$ on $M$ and a local coordinate
$B_i$ of the fiber, we can define a Poisson bracket as follows:
%
\begin{eqnarray}
\{ F, G \} \equiv 
F \frac{\rd}{\partial \phi^i} 
\frac{\ld }{\partial B_i}  G
- 
F \frac{\rd }{\partial B_i} 
\frac{\ld }{\partial \phi^i} G,
\label{ab0ike}
\end{eqnarray}
where $F$ and $G$ are functions on $T^* M$, and
${\rd}/{\partial \varphi}$ and ${\ld}/{\partial
\varphi}$ are the right and left differentiations
with respect to $\varphi$, respectively.
%
Here we shift the degree of fiber by $p$, 
i.e. the space $T^* [p]M$.
Then a Poisson structure shifts to a graded Poisson structure.
The corresponding graded Poisson bracket is called {\it antibracket}, 
$\sbv{*}{*}$.
%
Let $\bphi^i$ be a local coordinate of $M$
 and $\bb_{n-1,i}$ a basis of the fiber of $T^* [p] M$.
The antibracket $\sbv{*}{*}$ 
on a cotangent bundle $T^* [p] M$ is expressed as:
%
\begin{eqnarray}
\sbv{F}{G} \equiv 
F \frac{\rd}{\partial \bphi^i} 
\frac{\ld }{\partial \bb_{p,i}}  G
- 
F \frac{\rd }{\partial \bb_{p,i}} 
\frac{\ld }{\partial \bphi^i} G.
\label{Poisson2ike}
\end{eqnarray}
The total degree of the antibracket $\sbv{*}{*}$ is $-p$.
This antibracket satisfies the property (\ref{BVidentity})
for $-p = -n +1$.


Next, we consider a vector bundle $E \oplus E^*$. 
There is a natural Poisson structure on the fiber of $E \oplus E^*$ induced 
from a paring of $E$ and $E^*$.
If we take a local coordinate $A^a$ on the fiber of $E$ and 
$B_a$ on the fiber of $E^*$, we can define
\begin{eqnarray}
\{ F, G \} \equiv 
F \frac{\rd}{\partial A^a} 
\frac{\ld }{\partial B_a}  G
- 
F \frac{\rd }{\partial B_a} 
\frac{\ld }{\partial A^a} G,
\label{abpike}
\end{eqnarray}
where $F$ and $G$ are functions on $E \oplus E^*$.
We shift the degrees of fibers of $E$ and $E^*$ like $E [p] \oplus E^*[q]$, where
$p$ and $q$ are positive integers.
The Poisson structure changes to a graded Poisson structure
$\sbv{*}{*}$.
Let $\ba_p{}^{a}$ be a basis of the fiber of 
$E[p]$ and 
$\bb_{q,a}$ a basis of the fiber of 
$E^*[q]$.
The antibracket is represented as
\begin{eqnarray}
\sbv{F}{G} \equiv 
F  \frac{\rd}{\partial \ba_p{}^{a}} 
\frac{\ld }{\partial \bb_{q,a}}  G
- (-1)^{p q}
F \frac{\rd }{\partial \bb_{q,a}} 
\frac{\ld }{\partial \ba_p{}^{a}}  G.
\label{Poisson4ike}
\end{eqnarray}
The total degree of the antibracket $\sbv{*}{*}$ is $-p-q$.
This antibracket satisfies the property (\ref{BVidentity})
for $-p-q = -n +1$.


We define a {\it Q-structure}.
A {\it Q-structure} is 
a function $S$ on a graded manifold $\tilde{N}$
which satisfies the classical master equation 
$\sbv{S}{S} = 0$.
$S$ is called a {\it Batalin-Vilkovisky action}.  
We require that $S$ satisfy the compatibility condition
\begin{eqnarray}
S \sbv{F}{G} = \sbv{S F}{G} + (-1)^{|F| +1} \sbv{F}{SG},
\end{eqnarray}
where $F$ and $G$ are arbitrary functions.
$\sbv{S}{F} = \brs F$ generates an infinitesimal transformation, 
which is a {\it BRST transformation}, which
coincides with the gauge transformation of the theory.

The AKSZ formulation of the Batalin-Vilkovisky formalism is defined as 
a {\it P-structure} and a {\it Q-structure} on a {\it graded manifold}.


\subsection*{B-2. Batalin-Vilkovisky Structures of Topological Sigma Models}
\noindent
%
%
%
In this subsection, we explain Batalin-Vilkovisky structures 
of topological sigma models.
Let $X$ be a base manifold in $n$ dimensions, with or without boundary, 
and $M$ be a target manifold in $d$ dimensions.
We denote $\phi$ a smooth map from $X$ to $M$.
%

We consider a {\it supermanifold} $\Pi T X$, whose bosonic part is $X$.
$\Pi T X$ is defined as 
a tangent bundle with reversed parity of the fiber.
We take a local coordinate of $\Pi T X$, $(\sigma^{\mu}, \theta^{\mu})$,
where $\sigma^{\mu}$ is a coordinate on the base space and
$\theta^{\mu}$ is a super coordinate on the fiber and 
$\mu = 1, 2, \cdots, n$. 
We extend a smooth function $\phi$ to a function on the 
supermanifold $\bphi:\Pi T X \rightarrow M$.
$\bphi$ is called a superfield and an element of $\Pi T^* X \otimes M$.
We introduce a new non-negative integer grading on $\Pi T^* X$.
A coordinate $\sigma^{\mu}$ on a base manifold has zero 
and a coordinate $\theta^{\mu}$ on the fiber has one.
This grading is called the {\it form degree}.
We denote ${\rm deg} F$ the form degree of the function $F$.
The {\it total degree} defined in the previous section is a grading 
with respect to $M$, on the other hand 
The {\it form degree} is a grading 
with respect to $X$.
We define a {\it ghost number} ${\rm gh} F$ such that
${\rm gh} F = |F| - {\rm \deg} F$.
W assign the ghost numbers of $\sigma^{\mu}$ and $\theta^{\mu}$ zero.
Thus $\sigma^{\mu}$ has the total degree zero and 
$\theta^{\mu}$ has total degree one.

We consider a {\it P-structure} on $T^* [p] M$.
We take $p=n-1$ to construct
a Batalin-Vilkovisky structure in a topological sigma model
on a general $n$ dimensional worldvolume.
We consider $T^* [n-1] M$ for an $n$-dimensional base manifold $X$.
Let a superfield $\bphi^i$ be local a coordinate of $\Pi T^* X \otimes M$, 
where $i, j, k, \cdots$ are indices of the local coordinate on $M$.
Let a superfield $\bb_{n-1,i}$ be a basis 
of sections of $\Pi T^* X \otimes \bphi^*(T^* [n-1] M)$.
Expansions to component fields of the superfields are the following:
\begin{eqnarray}
&& \bphi^i = \phi^{(0)i} + \theta^{\mu_1} \phi_{\mu_1}^{(-1)i}
+ \frac{1}{2!} \theta^{\mu_1} \theta^{\mu_2} \phi_{{\mu_1}{\mu_2}}^{(-2)i}
+ \cdots + \frac{1}{n!} \theta^{\mu_1} \cdots \theta^{\mu_n} 
\phi_{{\mu_1} \cdots {\mu_n}}^{(-n)i},
\\ 
&& 
\bb_{n-1,i} = B_{n-1,i}^{(n-1)} 
+ \cdots 
+ \frac{1}{(n-1)!} \theta^{\mu_1} \cdots \theta^{\mu_{n-1}}
B_{{\mu_1} \cdots {\mu_{n-1}} n-1,i}^{(0)}
+ \frac{1}{n!} \theta^{\mu_1} \cdots \theta^{\mu_n}
B_{{\mu_1} \cdots {\mu_n} n-1,i}^{(-1)},
\nonumber 
\end{eqnarray}
where $(p)$ is the {\it ghost number} of the component field.

From (\ref{Poisson2ike}) in the previous subsection, 
we define an {\it antibracket} $\sbv{*}{*}$ 
on a cotangent bundle $T^* [n-1] M$ as
%
\begin{eqnarray}
\sbv{F}{G} \equiv 
F \frac{\rd}{\partial \bphi^i} 
\frac{\ld }{\partial \bb_{n-1,i}}  G
- 
F \frac{\rd }{\partial \bb_{n-1,i}} 
\frac{\ld }{\partial \bphi^i} G,
\label{cotangentPike}
\end{eqnarray}
where 
$F$ and $G$ are functions of $\bphi^i$ and $\bb_{n-1,i}$.
The total degree of the antibracket is $-n+1$.
If $F$ and $G$ are functionals of 
$\bphi^i$ and $\bb_{n-1,i}$, 
we understand an antibracket is defined as 
\begin{eqnarray}
\sbv{F}{G} \equiv 
\int_{\Pi T X} 
F \frac{\rd}{\partial \bphi^i} 
\frac{\ld }{\partial \bb_{n-1,i}}  G
- 
F \frac{\rd }{\partial \bb_{n-1,i}} 
\frac{\ld }{\partial \bphi^i} G,
\end{eqnarray}
where the integration $\int_{\Pi T X}$ means
the integration on the supermanifold, 
$\int_{\Pi T X} d^n \theta d^n \sigma$.
Through this article, 
we always understand an antibracket on two functionals in a similar 
manner and abbreviate this notation.

Next we consider a {\it P-structure} on $E \oplus E^*$.
In a topological sigma model in $n$ dimension worldvolume, 
we assign the total degree of 
$p$ and $q$ such that $p+q = n-1$.  
The total graded bundle is $E[p] \oplus E^*[n-p-1]$, 
where $-n+1 \leq p \leq n-1, p \neq 0$. 
%
%
%
Let $\ba_p{}^{a_p}$ be a basis of sections of 
$\Pi T^* X \otimes \bphi^*(E[p])$ and 
$\bb_{n-p-1,a_p}$ a basis of the fiber of 
$\Pi T^* X \otimes \bphi^*(E^*[n-p-1])$.
Expansions to component fields of the superfields are 
\begin{eqnarray}
&& \ba_p{}^{a_p} = A_p^{(p)a_p}
+ \theta^{\mu_1} A_{{\mu_1} p}^{(p-1)a_p} 
+ \cdots + 
+ \frac{1}{(p-1)!} \theta^{\mu_1} \cdots \theta^{\mu_(p-1)}
A_{{\mu_1} \cdots {\mu_(p-1)} p}^{(0)a_p}
\nonumber 
\\ 
&& \qquad 
+ \cdots 
+ \frac{1}{n!} \theta^{\mu_1} \cdots \theta^{\mu_n},
 A_{{\mu_1} \cdots {\mu_n} p}^{(-n+p)a_p} 
\\ 
&& 
\bb_{n-p-1,a_p} = 
B_{n-p-1,a_p}^{(n-p-1)}
+ \theta^{\mu_1} B_{{\mu_1} n-p-1,a_p}^{(n-p-2)}
+ \cdots
+ \frac{1}{(n-p-1)!} \theta^{\mu_1} \cdots \theta^{\mu_{(n-p-1)}}
B_{{\mu_1} \cdots {\mu_(n-p-1)} n-p-1,a_p}^{(0)}
\nonumber 
\\ 
&& \qquad 
+ \cdots 
+ \frac{1}{n!} \theta^{\mu_1} \cdots \theta^{\mu_n}
B_{{\mu_1} \cdots {\mu_n} n-p-1,a_p}^{(-p-1)},
\nonumber 
\end{eqnarray}
From (\ref{Poisson4ike}), we define the antibracket 
as
\begin{eqnarray}
\sbv{F}{G} \equiv 
F  \frac{\rd}{\partial \ba_p{}^{a_p}} 
\frac{\ld }{\partial \bb_{n-p-1,a_p}}  G
- (-1)^{n p}
F \frac{\rd }{\partial \bb_{n-p-1,a_p}} 
\frac{\ld }{\partial \ba_p{}^{a_p}}  G.
\label{EEPike}
\end{eqnarray}


We need to consider various grading assignments for $E \oplus E^*$,
because each assignment 
induces different Batalin-Vilkovisky structures.
In order to consider all independent assignments, 
we define the following bundle.
Let $E_p$ be 
series of vector bundles, 
where 
$-n+1 \leq p \leq n-1$.  
We consider a direct sum  of each bundle
$E_p[p]$ :
%
\begin{eqnarray}
\sum_{p=-n+1, p \neq 0}^{n-1}
E_p[p], 
\label{totbundleike}
\end{eqnarray}
%
and we can define a {\it P-structure}
on the graded vector bundle
\begin{eqnarray}
T^*[n-1] M \oplus 
\left(
\sum_{p=-n+1, p \neq 0}^{n-1}
E_p[p] \oplus E_p^*[n-p-1] \right),
\label{totspaceike}
\end{eqnarray}
%
which is isomorphic to the graded bundle
\begin{eqnarray}
T^*[n-1] \left(
\sum_{p=-n+1, p \neq 0}^{n-1}
E_p[p] \right).
\label{totspace2ike}
\end{eqnarray}
%
%
as a sum of (\ref{cotangentPike}) and (\ref{EEPike}):
\begin{eqnarray}
\sbv{F}{G} \equiv 
\sum_{p=-n+1}^{n-1}
F \frac{\rd}{\partial \ba_p{}^{a_p}} 
\frac{\ld }{\partial \bb_{n-p-1 \ a_p}} G
- (-1)^{n p}
F \frac{\rd }{\partial \bb_{n-p-1 \ a_p}} 
\frac{\ld }{\partial \ba_p{}^{a_p}} G.
\label{bfantibracketike}
\end{eqnarray}
where $\ba_0{}^{a_0} = \bphi^i$, that is
$p=0$ component is the antibracket (\ref{cotangentPike}) on 
the graded cotangent bundle $T^* [n-1] M$.
Note that all terms of the antibracket have 
the total degree $-n+1$, and 
we can confirm that 
the antibracket (\ref{bfantibracketike}) satisfies the identity
(\ref{BVidentity}).


\newcommand{\bibit}{\sl}


\vfill\eject
\end{document}